\documentclass[10pt]{elsarticle}
\usepackage[utf8]{inputenc}
\usepackage{amssymb,amsmath,array, algorithmic, algorithm}
\usepackage{xcolor}
\usepackage[toc,page]{appendix}

\usepackage{subcaption}
\usepackage{placeins}
\usepackage{hyperref}

\usepackage[all]{nowidow}
\usepackage{enumitem}

\usepackage{natbib}
\usepackage{pifont}% http://ctan.org/pkg/pifont
\newcommand{\cmark}{\ding{51}}%
\newcommand{\xmark}{\ding{55}}%

\newcommand{\setS}{\mathcal{S}}
\newcommand{\setX}{\mathcal{X}}
\newcommand{\setY}{\mathcal{Y}}
\newcommand{\setZ}{\mathcal{Z}}
\newcommand{\trainsetX}{\mathcal{D}_X}
\newcommand{\trainsetY}{\mathcal{D}_Y}
\newcommand{\vect}{\text{vect}}

\begin{document}
\title{Pixel-wise Conditioned Generative Adversarial Networks for Image Synthesis and Completion}

%***********************************************************************
% AUTHORS INFORMATION AREA
%***********************************************************************

\author{Cyprien Ruffino$^1$, Romain Hérault$^1$, Eric Laloy$^2$, Gilles Gasso$^1$
%
% Optional short acknowledgment: remove next line if non-needed
%
% DO NOT MODIFY THE FOLLOWING '\vspace' ARGUMENT
\vspace{.3cm}\\
%
% Addresses and institutions (remove "1- " in case of a single institution)
1- Normandie Univ, UNIROUEN, UNIHAVRE, INSA Rouen, LITIS\\ 76~000 Rouen, France
%
% Remove the next three lines in case of a single institution
\vspace{.1cm}\\
2- Belgian Nuclear Research, Institute Environment, Health and Safety,\\ Boeretang 200 - BE-2400 Mol, Belgium
}

\begin{abstract}
    Generative Adversarial Networks (GANs) have proven successful for unsupervised image generation. Several works have extended GANs to image inpainting by conditioning the generation with parts of the image to be reconstructed. Despite their success, these methods have limitations in settings where only a small subset of the image pixels is known beforehand. In this paper we investigate the effectiveness of conditioning GANs when very few pixel values are provided. We propose a modelling framework which results in adding an explicit cost term to the GAN objective function to enforce pixel-wise conditioning. We investigate the influence of this regularization term on the quality of the generated images and the fulfillment of the given pixel constraints. Using the recent PacGAN technique, we ensure that we keep diversity in the generated samples. Conducted experiments on FashionMNIST show that the regularization term effectively controls the trade-off between quality of the generated images and the conditioning. Experimental evaluation on the CIFAR-10 and CelebA datasets evidences that our method achieves accurate results both visually and quantitatively in term of Fréchet Inception Distance, while still enforcing the pixel conditioning. We also evaluate our method on a texture image generation task using fully-convolutional networks. As a final contribution, we apply the method to a classical geological simulation application.
\end{abstract}
\maketitle

\section{Introduction}

    Generative modelling is the process of modelling a distribution in a high-dimension space in a way that allows sampling in it. Generative Adversarial Networks (GANs) \cite{Goodfellow2014} have been the state of the art in unsupervised image generation for the past few years, being able to produce realistic images with high resolution \cite{Brock2018} without explicitly modelling the samples distribution. GANs learn a mapping function of vectors drawn from a low dimensional latent distribution (usually normal or uniform) to high dimensional ground truth images issued from an unknown and complex distribution. By using a discrimination function that distinguishes real images from generated ones, GANs setups a min max game able to approximate a Jensen-Shannon divergence between the distributions of the real samples and the generated ones.

    Among extensions of GANs, Conditional GAN (CGAN) \cite{mirza2014}  attempts to condition the generation procedure on some supplementary information $y$ (such as the label of the image $x$) by providing $y$ to the generation and discrimination functions. CGAN enables a variety of conditioned generation, such as class-conditioned image generation \cite{mirza2014}, image-to-image translation \cite{Isola2017, wang2018high}, or image inpainting \cite{pathak2016context}. On the other side, Ambient GAN \cite{bora2018ambientgan} aims at training an unconditional generative model using only noisy or incomplete samples $y$. Relevant application domain is high-resolution imaging (CT scan, fMRI) where image sensing may be costly. Ambient GAN attempts to produce unaltered images $\tilde{x}$ which distribution matches the true one without accessing to the original images $x$. For the sake, Ambient GAN considers lossy measurements such as blurred images, images with removed patch or removed pixels at random (up to 95\%). Following this setup, Pajot et al.\cite{pajot2018unsupervised} extend the learning strategy to enable the reconstruction instead of the generation of realistic images from similarly altered samples. 
    
    In the spirit of Ambient GAN,  we consider in this paper an extreme setting of image generation when only a few pixels, less than a percent of the  image size, are known and are randomly scattered across the image (see Fig.\ref{fig:pixelwise_gen}). We refer to these conditioning pixels as a constraint map $y$. To reconstruct the missing information, we design a generative adversarial model able to generate high quality images coherent with given pixel values by leveraging on a training set of similar, but not paired images. The model we propose aims to match the distribution of the real images conditioned on a highly scarce constraint map, drawing connections with Ambient GAN while, in the same manner as CGAN, still allowing the generation of diverse samples following the underlying conditional distribution. 
    
    To make the generated images honoring the prescribed pixel values, we use a reconstruction loss measuring how close real constrained pixels are to their generated counterparts. We show that minimizing this loss is equivalent to maximizing the log-likelihood of the constraints given the generated image. Thereon we derive an objective function trading-off the adversarial loss of GAN and the reconstruction loss which acts as a regularization term. We analyze the influence of the related hyper-parameter in terms of quality of generated images and the respect of the constraints. Specifically, empirical evaluation on FashionMNIST~\cite{Xiao2017} evidences that the regularization parameter allows for controlling the trade-off between samples quality and constraints fulfillment.
    
    Additionally to show the effectiveness of our approach, we conduct experiments on CIFAR10 \cite{Krizhevsky2009CIFAR10}, CelebA \cite{liu2015celeba} or texture \cite{jetchev2016texture} datasets using various deep architectures including fully convolutional network. We also evaluate our method on a classical geological problem which consists of generating 2D geological images of which the spatial patterns are consistent with those found in a conceptual image of a binary fluvial aquifer\cite{Strebelle2002}\cite{laloy2018training}. Empirical findings reveal that the used architectures may lack stochasticity from the generated samples that is the GAN input is often mapped to the same output image irrespective of the variations in latent code \cite{yang2018diversitysensitive}. We address this issue by resorting to the recent PacGAN \cite{lin2018pacgan} strategy.
    As a conclusion, our approach performs well both in terms of visual quality and respect of the pixel constraints while keeping diversity among generated samples. Evaluations on CIFAR-10 and CelebA show that the proposed generative model always outperforms the CGAN approach on the respect of the constraints and either come close or outperforms it on the visual quality of the generated samples.
    
    The remainder of the paper is organized as follows. In Section \ref{sec:related_work}, we review the relevant related  work focusing first on generative adversarial networks, their conditioned version and then on methods dealing with image generation and reconstruction from highly altered training samples.  Section \ref{sec:our-approach}  details the overall generative model we propose. In Section \ref{sec:experiments_protocol}, we present the experimental protocol and evaluation measures while Section \ref{sec:results} gathers quantitative and qualitative effectiveness of our approach. The last section concludes the paper.
    
    The contributions of the paper are summarized as follows:
    \begin{itemize}[nosep]
        \item We propose a method for learning to generate images with a few pixel-wise constraints.
        \item A theoretical justification of the modelling framework is investigated.
        \item A controllable trade-off between the image quality and the constraints' fulfillment is highlighted,
        \item We showcase a lack of diversity in generating high-dimensional images which we solve by using  PacGAN\cite{lin2018pacgan} technique. Several experiments allow to conclude that the proposed formulation can effectively generate diverse and high visual quality images while satisfying the pixel-wise constraints. 
    \end{itemize}

\section{Image reconstruction with GAN in related works}\label{sec:related_work}

       The pursued objective of the paper is image generation using generative deep network conditioned on  randomly scattered and scarce (less than a percent of the image size) pixel values. This kind of pixel constraints occurs in application domains where an image or signal need to be generated from very sparse measurements.
        
        Before delving into the details, let introduce the notations and previous work related to the problem. We denote by $X \in \setX$ a random variable and $x$ its realization. Let $p_X$ be the distribution of $X$ over $\setX$ and $p_X(x)$ be its evaluation at $x$. Similarly $p_{X|Y}$ represents the distribution of $X$ conditioned on the random variable $Y \in \setY$. 

\begin{figure}[t]
		    \centering
		    \begin{subfigure}[t]{0.33\textwidth}
		        \centering
		        \includegraphics[width=3cm]{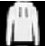}
		        \caption{Original \\ Image}
                \label{fig:digit}
            \end{subfigure}\begin{subfigure}[t]{0.33\textwidth}
		        \centering
		        \includegraphics[width=3cm]{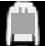}
		        \caption{Inpainting\\Input}
                \label{fig:inpainting}
		    \end{subfigure}\begin{subfigure}[t]{0.33\textwidth}
		        \centering
		        \includegraphics[width=3cm]{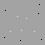}
		        \caption{Constraint\\Map}
                \label{fig:pixelwise_gen}
		    \end{subfigure}
		    \caption{Difference between regular inpainting (\ref{fig:inpainting}) and the problem undertaken in this work (\ref{fig:pixelwise_gen}) on a real sample (\ref{fig:digit}).}
		    \label{fig:image_completion_task}
		\end{figure}

\begin{figure}[t]
		    \centering
		    \begin{subfigure}[t]{0.25\textwidth}
		        \centering
		        \includegraphics[scale=1.5]{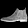}
		        \caption{Original\\Image}
                \label{fig:original_shoe}
            \end{subfigure}\begin{subfigure}[t]{0.25\textwidth}
		        \centering
		        \includegraphics[scale=1.5]{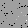}
		        \caption{Constraints}
                \label{fig:constraints}
		    \end{subfigure}\begin{subfigure}[t]{0.25\textwidth}
		        \centering
		        \includegraphics[scale=1.5]{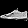}
		        \caption{Generated\\Image}
                \label{fig:pixelwise}
		     \end{subfigure}\begin{subfigure}[t]{0.24\textwidth}
		        \centering
		        \includegraphics[scale=1.5]{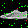}
		        \caption{Satisfied\\Consts.}
                \label{fig:generated}
		    \end{subfigure}
		    \caption[Generation of a sample during training]{Generation of a sample during training. We first sample an image from a training set (\ref{fig:original_shoe}) and we sample the constraints (\ref{fig:constraints}) from it. Then our GAN generates a sample (\ref{fig:pixelwise}). The constraints with squared error smaller than $\epsilon=0.1$ are deemed satisfied and shown by green pixels in (\ref{fig:generated}) while the red pixels are unsatisfied.}
		    \label{fig:image_completion}
		\end{figure}

        Given a set of images $x \in \setX = [-1, 1]^{n\times p\times c}$  (see Figure \ref{fig:digit}) drawn from an unknown distribution $p_X$ and a sparse matrix  $y \in  \setY = [-1, 1]^{n\times p\times c}$ (Figure \ref{fig:pixelwise_gen}) as the given constrained pixels, the problem consists in finding a generative model $G$ with inputs $z$ (a random vector sampled from a known distribution $p_Z$ over the space $\setZ$) and constrained pixel values $y \in  [-1, 1]^{n\times p\times c}$ able to generate an image satisfying the constraints while likely following the distribution $p_X$ (see Figure \ref{fig:image_completion}).

\begin{figure}[h]
		    \centering
		    \begin{subfigure}[b]{0.45\textwidth}
		        \includegraphics[width=\textwidth]{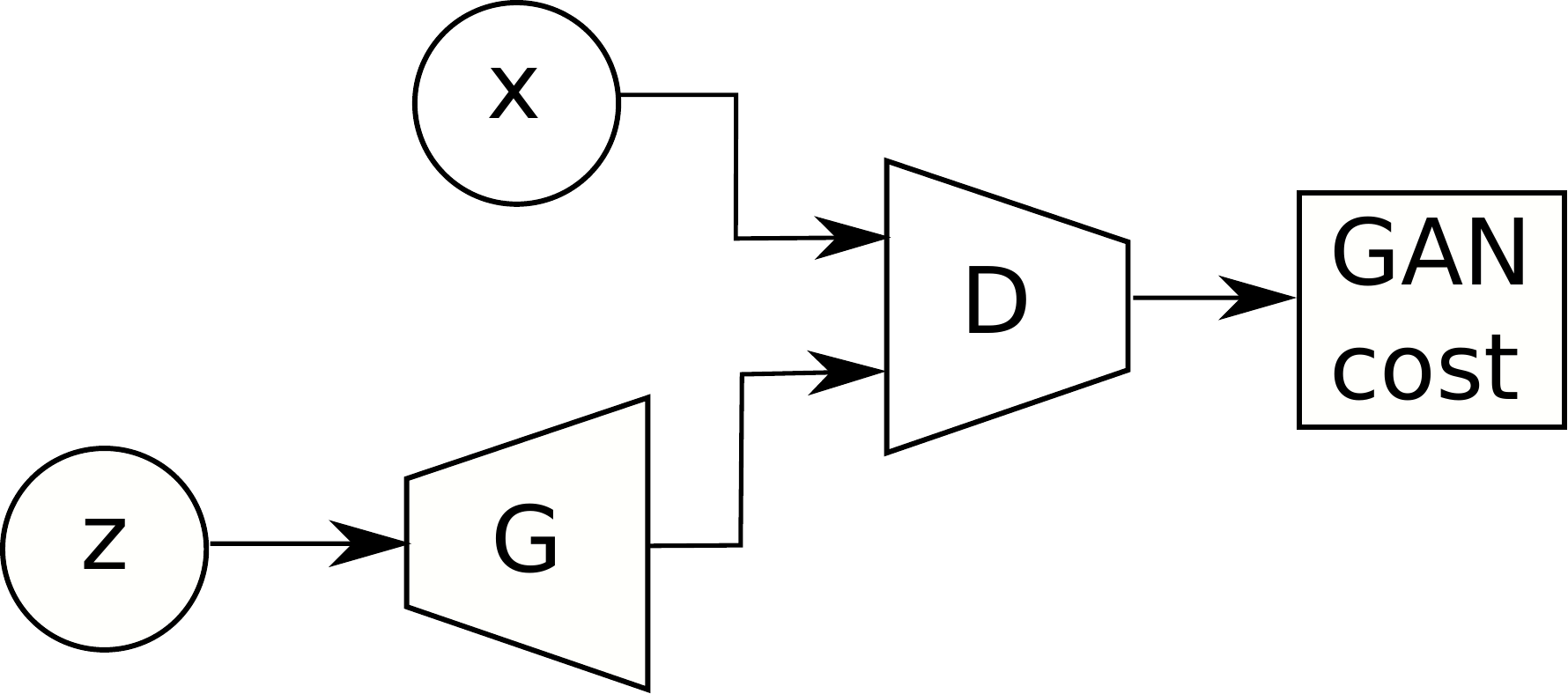}
		        \caption{GAN}
                \label{fig:gan}
		    \end{subfigure}
		    \begin{subfigure}[b]{0.45\textwidth}
		        \includegraphics[width=\textwidth]{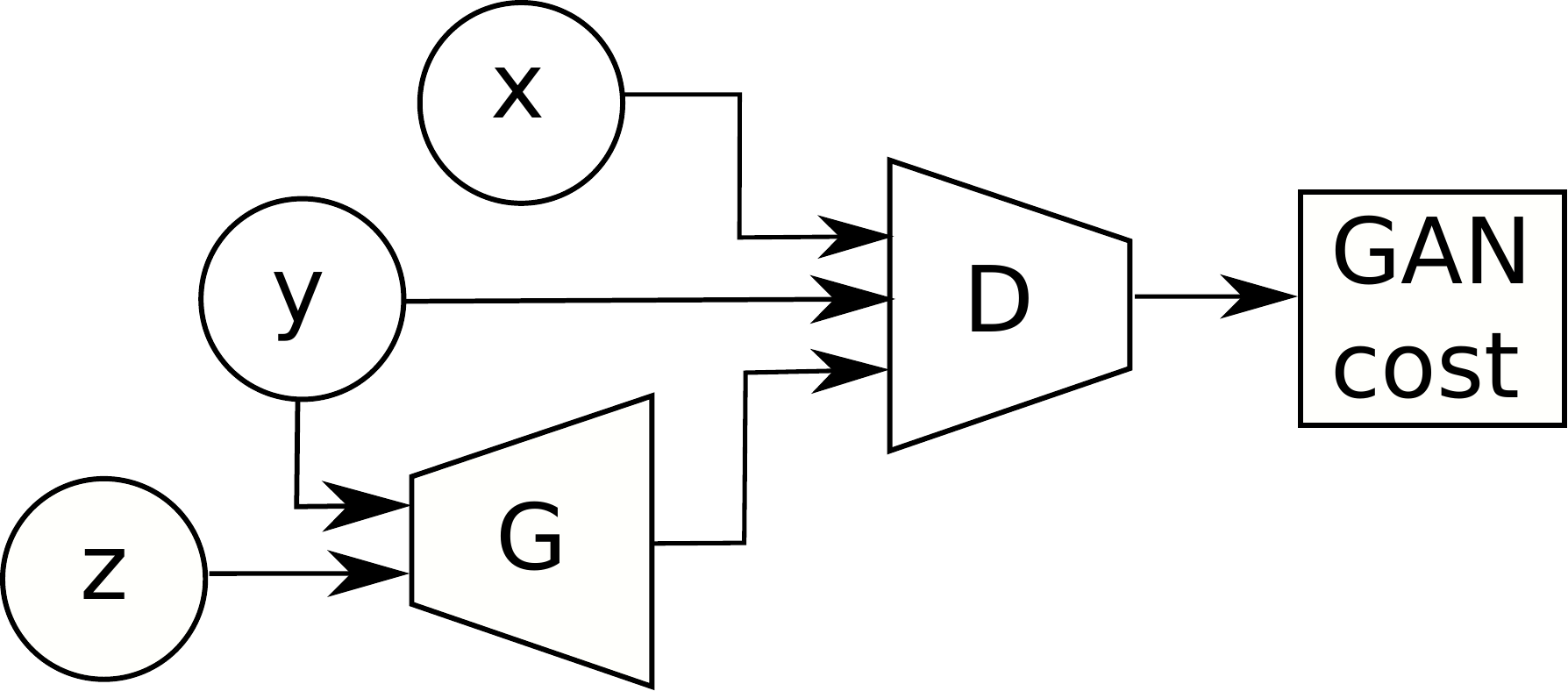}
		        \caption{CGAN}
                \label{fig:cgan}
		    \end{subfigure}
		    \begin{subfigure}[b]{0.45\textwidth}
		        \includegraphics[width=\textwidth]{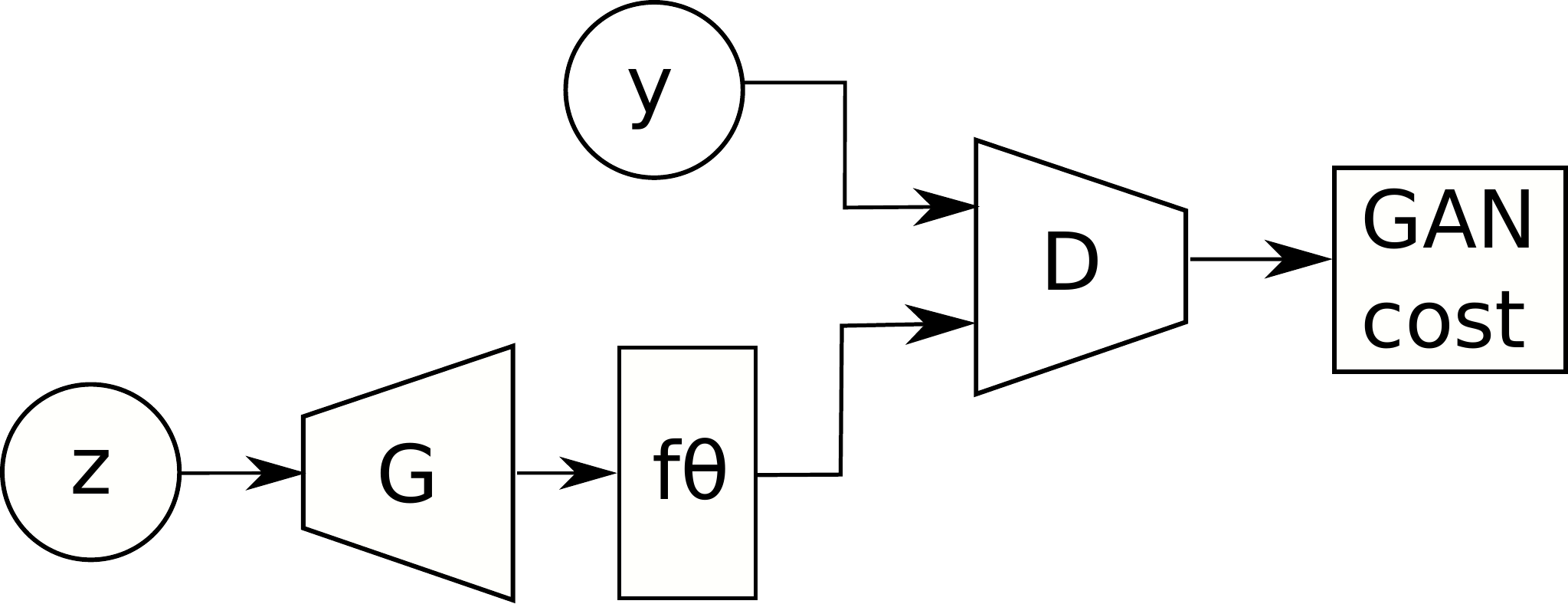}
		        \caption{AmbientGAN}
                \label{fig:ambientgan}
		    \end{subfigure}
		    \hspace{3mm}
		    \begin{subfigure}[b]{0.45\textwidth}
		        \includegraphics[width=\textwidth]{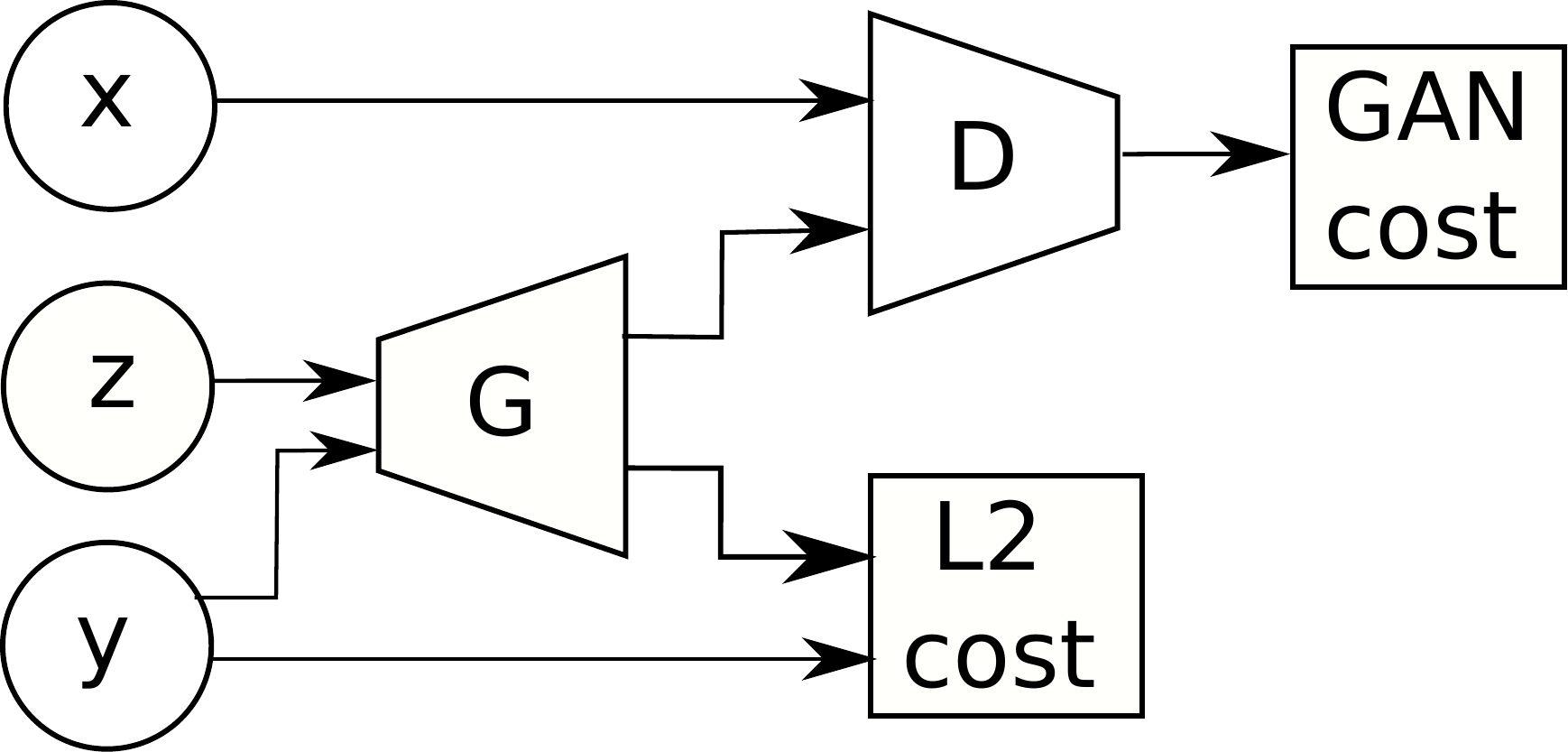}
		        \caption{Our approach}
                \label{fig:ourapproach}
		    \end{subfigure}
		    \caption{Different GAN Setups. G and D are the generator and discriminator networks, x and z are samples from the distributions $P_x$ and $P_r$, y is a label/constraint map sampled from $P_y$ and $f_\theta$ is an image degradation function.}
		    \label{fig:gansetup}
		\end{figure}

         One of the state-of-the-art modelling framework for image generation is the Generative Adversarial Network. The seminal version of GAN \cite{Goodfellow2014} learns the generative models in an unsupervised way.
		 It relies on a game between a generation function $G$ and a discrimination network $D$, in which $G$ learns to produce realistic samples while $D$ learns to distinguish real examples from generated ones (Figure \ref{fig:gan}). Training GANs amounts to find a Nash equilibrium to the following min-max problem,

		\begin{equation}
		    \label{eq:basic_gan}
			\min_G \max_D L(D, G) = \mathop{\mathbb{E}}_{x\sim p_X} \Big[\log (D(x))\Big] + \mathop{\mathbb{E}}_{z\sim p_Z} \Big[\log (1-D(G(z)))\Big] \enspace,
		\end{equation}

		\noindent where $p_Z$ is a known distribution, usually normal or uniform, from which the latent input $z$ of $G$ is drawn, and $p_X$ is the distribution of the real images. %However, this objective has been shown to be particularly unstable, so in practice a non-saturating version of this cost is used , as recommended in \cite{Goodfellow2014}.

        Among several applications, the GANs was adapted  to image inpainting task (Figure \ref{fig:inpainting}). For instance Yeh et al. \cite{Yeh2017} propose an inpainting approach which considers a pre-trained generator, and explores its latent space $\mathcal{Z}$ through an optimization procedure to find a latent vector $z$, which induces an image with missing regions filled in by conditioning on the surroundings available information. However, the method requires to solve a full optimization problem at inference stage, which is computationally expensive.
        
        Other approaches (Figure \ref{fig:gansetup}) rely on Conditional variant of GAN (CGAN) \cite{mirza2014} in which additional information $y$ is provided to the generator and the discriminator (see Figure \ref{fig:cgan}). 
        This leads to the following optimization problem adapted to CGAN
		%\vspace{-0.5em}
		\begin{equation}
		\min_G \max_D L(D, G) = \mathop{\mathbb{E}}_{\substack{x\sim p_X \\y \sim p_{Y|X} }} \Big[\log(D(x, y))\Big]
		+ \mathop{\mathbb{E}}_{\substack{z\sim p_Z\\y\sim p_Y}} \Big[ \log(1-D(G(y,z),y))\Big] \enspace.
		\end{equation}
        
        Although CGAN was initially designed for class-conditioned image generation by setting $y$ as the class label of the image, several types of conditioning information can apply such as a full image for image-to-image translation \cite{Isola2017} or partial image as in inpainting \cite{Yu2018}. CGAN-based inpainting methods rely on generating a patch that will fill up a structured missing part of the image and achieve impressive results. However they are not well suited to reconstruct very sparse and unstructured signal \cite{demir2018}. Additionally, these approaches learn to reconstruct a single sample instead of a full distribution, implying that there is no sampling process for a given constraint map or highly degraded image.

        AmbientGAN \cite{bora2018ambientgan} (Figure \ref{fig:ambientgan}) trains a generative model capable to yield full images from only lossy measurements. One of the image degradations considered in this approach is the random removal of pixels leading to sparse pixel map $y$. It is simulated with a differentiable function $f_\theta$ whose parameter $\theta$ indicates the pixels to be removed. The underlying optimization problem solved by AmbientGAN is therefore stated as
    	%\vspace{-0.5em} % pas de vspace sinon çà merde les titres
		\begin{equation}
	    	\min_G \max_D L(D, G) = \mathop{\mathbb{E}}_{\substack{y\sim p_Y}} \Big[\log(D(y))\Big] + \mathop{\mathbb{E}}_{\substack{z\sim p_Z \\\theta \sim p_\theta}} \Big[ \log(1-D(f_\theta(G(z))))\Big] \enspace.
		\end{equation}
	
		Pajot et al. \cite{pajot2018unsupervised} combined the AmbientGAN approach with an additional reconstruction task that consists in reconstructing the $f_\theta(G(y))$ from the twice-altered image $\tilde{y} = f_\theta(G(y))$ and $\hat{y} = f_\theta(G(f_\theta(G(y))))$,
		
		\begin{equation}
	    	\min_G \max_D L(D, G) = \mathop{\mathbb{E}}_{\substack{y\sim p_Y}} \Big[\log(D(y))\Big] + \mathop{\mathbb{E}}_{\substack{y\sim p_Y}} \Big[ \log(1-D(\hat{y}))\Big] + ||\hat{y} - \tilde{y} ||^2_2 \enspace.
		\end{equation}
		\noindent
		 The $\ell_2$ norm term ensures that the generator is able to learn to revert $f_\theta$ i.e. to revert the alteration process on a given sample. This  allows the reconstruction of realistic image only from a given constraint map $y$. However the reconstruction process is deterministic and does not provide a sampling mechanism.
		 
		 Compressed Sensing with Meta-Learning \cite{wu2019deep} is an approach that combines the exploration of the latent space $\mathcal{Z}$ to recover images from lossy measurements with the enforcing of the Restricted Isometric Property \cite{candes2005decoding}, which states that for two samples $x_1,x_2 \sim p_X$, $$(1 - \alpha)||x_1 - x_2||_2^2 \leq ||f_\theta(x_1 - x_2)||_2^2 \leq (1 + \alpha) ||x_1 - x_2||_2^2$$ where $\alpha$ is a small constant.
		 It replaces the adversarial training of the generative model $G$ (Eq. \ref{eq:basic_gan}) by searching, for a given degraded image $y$, a vector $\hat{z}$ such that $\hat{y} = f_\theta(G(\hat{z}))$ minimizes the $\ell_2$ distance between $y$ and $\hat{y}$ while still enforcing the RIP. The overall problem induced by this approach can be formulated as:
		
% 		\begin{multline}
% 	    	\min_G L(G) = \mathop{\mathbb{E}}_{\substack{x\sim p_X\\y\sim p_Y\\z\sim p_Z}} \Big[\Big( (||f_\theta (x - G(z))||_2^2 - ||x - G(z)||_2^2)^2 + (||f_\theta (x - G(\hat{z}))||_2^2 - ||x - G(\hat{z})||_2^2)^2 \\
% 	    	+ (||f_\theta (G(z) - G(\hat{z}))||_2^2 - ||G(z) - G(\hat{z})||_2^2)^2 \Big) / 3
% 	    	+ ||y - f_\theta(G(\hat{z})) ||^2_2\Big]\\
% 	    	\text{where } \hat{z} = \min_z ||y - f_\theta(G(z)) ||^2\enspace.
% 		\end{multline}
        
        \begin{multline}
	    	\min_G L(G) = \mathop{\mathbb{E}}_{\substack{x\sim p_X\\y\sim p_Y\\z\sim p_Z}} \Big( \sum_{\substack{x_1, x_2 \in \setS\\x_1 \ne x_2}}(||f_\theta (x_1 - x_2)||_2^2 - ||x_1 - x_2||_2^2)^2 \Big) / 3
	    	+ ||y - f_\theta(G(\hat{z})) ||^2_2\\
	    	\text{where } \hat{z} = \min_z ||y - f_\theta(G(z)) ||^2\enspace.
		\end{multline}
		\noindent
		 
		 where $\setS$ contains the three samples $x, G(z), G(\hat{z})$. 
		 In practice, $\hat{z}$ is computed with gradient descent on $z$ by minimizing $||y - f_\theta(G(z)) ||^2$, and starting from a random $z \sim p_Z$. As a benefit, this approach may generate an image $\hat{x} = G(\hat{z})$ from a noisy information $y$ but at a high computation burden since it requires to solve an optimization problem (computing $\hat{z}$) at inference stage for generating an image.
		 
\section{Proposed approach} \label{sec:our-approach}

    Let introduce the formal formulation of the addressed problem. Assume $y$ is the given set of constrained pixel values. To ease the presentation, let consider $y$ as a $n\times p\times c$ image with only a few available pixels (less than $1\%$ of $n\times p\times c$). We will also encode the spatial location of these pixels using a corresponding binary mask $M(y) \in \{0,1 \}^{n\times p\times c}$.  We intend to learn a GAN whose generation network takes as input the constraint map $y$ and the sampled latent code $z \in \setZ$ and outputs a realistic image that fulfills the prescribed pixel values. Within this setup, the generative model can sample from the unknown distribution $p_X$ of the training images $\{x_1, \cdots, x_N\}$ while satisfying unseen pixel-wise constraints at training stage. Formally our proposed GAN can be formulated as
	\begin{eqnarray}
	\label{eq:formulation_our_primary_GAN}
		\min_G \max_D L(D,G){\small=}\mathop{\mathbb{E}}_{\substack{x\sim p_{x}}} \Big[\log(D(x))\Big]{\small+} \mathop{\mathbb{E}}_{\substack{z{\small\sim} p_Z\\y{\small\sim} p_{Y}}} \Big[ \log(1{\small -}D(G(y, z)))\Big] \enspace,  \\
		\text{s.t. } y = M(y) \odot G(y,z) \nonumber
	\end{eqnarray}
	
	\noindent where $\odot$ stands for the Hadamard (or point-wise) product and $M(y)$ for the mask, a sparse matrix with entries equal to one at constrained pixels location. 
	
	As the equality constraint in  Problem (\ref{eq:formulation_our_primary_GAN}) is difficult to enforce during training, we rather investigate a relaxed version of the problems.
	%
	%A first way to do that is to use the aforementioned CGAN\cite{mirza2014} method, however we focused on using a regularization-based approach.
	%To justify our choice of a regularization term, we assume that the errors on the constraints $\epsilon$ follow some common distribution. We then formulate the reconstruction of the constraints as,
	%\begin{equation}
	 %   y = f(G(y, z)) + \epsilon \enspace.
	%\label{eq:noisy_generation_primary_CGAN}
	%\end{equation}
	%
	Following Pajot et al. \cite{pajot2018unsupervised} we assume that the constraint map is obtained through a noisy measurement process
	\begin{equation}
	    y = f_M(x) + \varepsilon \enspace.
	\label{eq:noisy_generation_primary_CGAN}
	\end{equation}
	Here $f_M$ is the masking operator yielding to $y = M(y) \odot x$. Also the constrained pixels are randomly and independently selected. $\varepsilon$ represents an additive i.i.d noise corrupting the pixels. Therefore
    we can formulate the Maximum A Posteriori (MAP) estimation problem, which, given the constraint map $y$, consists in finding the most probable image $x^*$ following the posterior distribution $p_{X|Y}$,
    \begin{align}
        x^* &= \arg\max_x\log {p_{X|Y}}(x|y) \\
            &= \arg\max_x\log p_{Y|X}(y|x) + \log p_X(x) \enspace.
	    \label{eq:bayesian_formulation_our_primary_CGAN}
	\end{align}

     \noindent $p_{Y|X}(y|x)$ is the likelihood that the constrained pixels $y$ are issued from image $x$ while $p_X(x)$ represents the prior probability at $x$. Assuming that the generation network $G$ may sample the most probable image $G(y, z)$ complying with the given pixel values $y$, we get the following problem
        
    \begin{equation}
        G^* = \arg\max_G \mathop{\mathbb{E}}_{\substack{y\sim p_Y\\z\sim p_Z}} \log {p_{Y|X}}(y|G(y, z)
        ) + \log p_X(G(y, z)) \enspace.
        \label{eq:bayesian_formulation_our_primary_CGAN_G}
    \end{equation}

\noindent The first term in Problem (\ref{eq:bayesian_formulation_our_primary_CGAN_G}) measures the likelihood of the constraints given a generated image. Let rewrite Equation (\ref{eq:noisy_generation_primary_CGAN}) as $\vect(y) = \vect(f_M(x)) + \vect(\varepsilon)$ where $\vect(\cdot)$ is the vectorisation operator that consists in stacking the constrained pixels. Therefore, assuming $\vect(\varepsilon)$ is an i.i.d Gaussian noise with distribution $\mathcal{N}(0,\sigma^2 I)$, we achieve the expression of the conditional likelihood
\begin{equation}
    log {p_{Y|X}}(y|G(y, z)) \, \propto - \left \|\vect(y) - \vect(M(y) \odot G(y,z))\right\|_2^2 \enspace
\end{equation}
\noindent which evaluates the quadratic distance between the conditioning pixels and their predictions by $G$. In other words, using a matrix notation of  (\ref{eq:noisy_generation_primary_CGAN}), the likelihood of the constraints given a generated image equivalently writes
   
   \begin{equation}
    \log {p_{Y|X}}(y|G(y, z) \, \propto - \left \|y - M(y) \odot G(y,z)\right\|_F^2 \enspace.
    \end{equation}
    
   \noindent $\| A \|_F^2 $ represents the squared Frobenius norm of matrix $A$ that is the sum of its squared entries. 
    %
    %In this work, we assume that $\epsilon$ follows a normal distribution $\epsilon\sim\mathcal{N}[0,\sigma^2]$, but it is worth noting that any prior distribution with a close-form solution for maximum likelihood estimation (typically distribution from the exponential family) can be used.
    %In the case of the normal distribution, we can minimize our error term by using the $L_2$ norm on the constrained pixels.
    %    
    
    The second term in Problem (\ref{eq:bayesian_formulation_our_primary_CGAN_G}) is the likelihood of the generated image under the true but unknown data distribution $p_X$. Maximizing this term can be equivalently achieved by minimizing the distance between $p_X$ and the marginal distribution of the generated samples $G(y,z)$. This amounts to minimizing with respect to $G$, the GAN-like objective function $\mathop{\mathbb{E}}_{\substack{x\sim p_X}} \log(D(x)) + \mathop{\mathbb{E}}_{\substack{z\sim p_Z\\y\sim p_Y}} \log(1-D(G(y, z)))$  \cite{Goodfellow2014}. Putting altogether these elements, we can propose a relaxation of the hard constraint optimization problem (\ref{eq:formulation_our_primary_GAN}) (Figure \ref{fig:ourapproach}) as follows
    %minimizing the Jensen-Shannon divergence between the real data distribution and the distribution of the generated data \cite{Goodfellow2014}.
	
	%In our approach, we explicitly model the relaxation of the constraint by minimizing the $L_2$ norm between the constrained pixels and the generated values (see Fig.\ref{fig:ourapproach}).
	
	%The objective function, with $\lambda \geq 0$ an additional parameter, becomes:
	\begin{eqnarray}
		\min_G \max_D L(D,G) & {\small=} & \mathop{\mathbb{E}}_{\substack{x\sim p_X}} \Big[\log(D(x))\Big] \label{eq:final_optim_problem} \\
		&+&\mathop{\mathbb{E}}_{\substack{z\sim p_Z\\y\sim p_Y}} \Big[\log(1-D(G(y, z)))+\lambda \left\|y - M(y) \odot G(y,z)\right\|_F^2 \Big] \enspace . \nonumber
	\end{eqnarray}

\subsubsection*{Remarks:}
\begin{itemize}
    \item The assumption of Gaussian noise measurement leads us to explicitly turn the pixel value constraints into the  minimization of the $\ell_2$ norm between the real enforced pixel values and their generated counterparts (see Figure \ref{fig:ourapproach}).
    
    \item This additional term acts as a regularization over prescribed pixels by the mask $M(y)$. The trade-off between the distribution matching loss and the constraint enforcement is assessed by the regularization parameter $\lambda \geq 0$.
    
    \item It is worth noting that the noise $\varepsilon$ can be of any other distribution, according to the prior information, one may associate to the measurement process. We only require this distribution to admit a closed-form solution for the maximum likelihood estimation for optimization purpose. Typical choices are distributions from the exponential family \cite{brown1986fundamentals}.
    
\end{itemize}

      To solve Problem (\ref{eq:final_optim_problem}), we use the stochastic gradient descent method. The overall training procedure is detailed in Algorithm \ref{alg:train} and ends up when a maximal number of training epochs is attained. 
      
      When implementing this training procedure we experienced, at inference stage, a lack of diversity in the generated samples (see Figure \ref{fig:diversity_loss}) with deeper architectures, most notably the encoder-decoder architectures. This issue manifests itself through the fact that the learned generation network, given a constraint map $y$, outputs almost deterministic image  regardless the variations in the input $z$. The issue was also pointed out by Yang et al. \cite{yang2018diversitysensitive} as characteristic of CGANs.

        \begin{algorithm}[!ht]
        \caption{Proposed training algorithm}
        \label{alg:train}
        \begin{algorithmic}[H]
           \REQUIRE{ $\trainsetX$ the set of  unaltered images, $\trainsetY$ the set of constraint maps, $G$ the generation network, and $D$ the discrimination function}
            \REPEAT
                \STATE sample a mini-batch $\lbrace x_i \rbrace_{i=1}^m$ from $\trainsetX$\;
                \STATE sample a mini-batch $\lbrace y_i \rbrace_{i=1}^m$ from $\trainsetY$\;
                \STATE sample a mini-batch $\lbrace z_i \rbrace_{i=1}^m$ from distribution $p_Z$ \;
                \STATE update $D$ by stochastic gradient ascent of
                \STATE \ \ \ \ $ \sum_{i=1}^{m}\log(D(x_i)) + \log(1-D(G(y_i, z_i)))$
                \STATE sample a mini-batch $\lbrace y_j \rbrace_{j=1}^n$ from $\trainsetY$\;
                \STATE sample a a mini-batch $\lbrace z_j \rbrace_{j=1}^n$ from distribution $p_Z$\;; 
                \STATE update $G$ by stochastic gradient descent of
                \STATE \ \ \ \ $ \sum_{j=1}^n \log(1-D(G(y_j, z_j))) + ||y_j - M(y_j)\odot G(y_j, z_j)||_F^2$\;
            \UNTIL a stopping condition is met
            
        \end{algorithmic}
        \end{algorithm}
        
    To avoid the problem, we exploit the recent PacGAN \cite{lin2018pacgan} technique: it consists in passing a set of samples to the discrimination function instead of a single one.  PacGAN is intended to tackle the mode collapse problem in GAN training. The underlying principle being that if a set of images are sampled from the same training set, they are very likely to be completely different, whereas if the generator experiences mode collapse, generated images are likely to be similar.
        In practice, we only give two samples to the discriminator, which is sufficient to overcome the loss of diversity as  suggested in \cite{lin2018pacgan}. 
        The resulting training procedure is summarized in Algorithm~\ref{alg:trainpac}.
        
         \begin{algorithm}[!ht]
        \caption{Our training algorithm including PacGAN}
        \label{alg:trainpac}
        \begin{algorithmic}[H]
           \REQUIRE { $\trainsetX$ the set of  unaltered images, $\trainsetY$ the set of constraint maps, $G$ the generation network, and $D$ the discrimination function}
            \REPEAT
                \STATE sample two mini-batches $\lbrace x_i^a \rbrace_{i=1}^m$, $\lbrace x_i^b\rbrace_{i=1}^m$ from $\trainsetX$\;
                \STATE sample a mini-batch $\lbrace y_i \rbrace_{i=1}^m$ from $\trainsetY$\;
                \STATE sample two mini-batches $\lbrace z_i^a \rbrace_{i=1}^m$, $\lbrace z_i^b \rbrace_{i=1}^m$ from distribution $p_Z$ \;
                \STATE update $D$ by stochastic gradient ascent of
                \STATE \ \ \ \ $ \sum_{i=1}^{m}\log(D(x_i^a, x_i^b)) + \log(1-D(G(y_i, z_i^a), G(y_i, z_i^b)))$
                \STATE sample a mini-batch $\lbrace y_j \rbrace_{j=1}^n$ from $\trainsetY$\;
                \STATE sample two mini-batches $\lbrace z_i^a \rbrace_{i=1}^m$, $\lbrace z_i^b \rbrace_{i=1}^m$ from distribution $p_Z$ \;
                \STATE update $G$ by stochastic gradient descent of
                \STATE \ \ \ \ $ \sum_{j=1}^n \log(1-D(G(y_j, z_j))) + ||y_j - M(y_j)\odot G(y_j, z_j)||_F^2$\;
            \UNTIL a stopping condition is met
            
        \end{algorithmic}
        \end{algorithm}

\FloatBarrier
		
\section{Experiments} \label{sec:experiments_protocol}
    We have conducted a series of empirical evaluation to assess the performances of the proposed GAN. Used datasets, evaluation protocol and the tested deep architectures are detailed in this section while Section \ref{sec:results} is devoted to the results presentation. 
	\subsection{Datasets}

        We tested our approach on several datasets listed hereafter. Detailed  information on these datasets are provided  in the Appendix \ref{app:det_datasets}.
        %namely FashionMNIST \cite{Xiao2017}, CIFAR10 \cite{Krizhevsky2009CIFAR10}, CelebA\cite{liu2015celeba} and a custom-made Texture texture dataset:
	    \begin{description}
	    \item{FashionMNIST} \cite{Xiao2017} consists of 60,000 $28\times 28$ small grayscale images of fashion items, split in 10 classes and is a harder version of the classical MNIST  dataset \cite{lecun1998}. %known to be simple to solve. 
	    The very small size of the images makes them particularly appropriate for large-scale experiments, such as hyper-parameter tuning. 
	    
	    \item{CIFAR10} \cite{Krizhevsky2009CIFAR10} consists of 60,000 $32 \times 32$ colour images of 10 different and varied classes. It is deemed less easy than MNIST and FashionMnist
	    %considered harder to learn that MNIST and FashionMNIST, even it is of nearly the same dimension.
	     \item{CelebA}\cite{liu2015celeba} is a large dataset of celebrity portraits labeled by identity and a variety of binary features such as eyeglasses, smiling... We use 100,000 images cropped to a size of $128 \times 128$, making this dataset appropriate for a high dimension evaluation of our approach in comparison with related work.

        \item{Texture} is a custom dataset 
        %was eventually created that is composed of texture, sampling $20000$ patches
        composed of $20,000$ $160 \times 160$ patches sampled from a large brick wall texture, as recommended in \cite{jetchev2016texture}. It is worth noting that this procedure can be reproduced on any texture image of sufficient size. Texture is a testbed of our approach on fully-convolutional networks for constrained texture generation task. 
        %This allows us to experiment fully-convolutional architectures on a texture reconstruction task.
        
        \item{Subsurface} is a classical dataset in geological simulation \cite{Strebelle2002} which consists, similarly to the Texture dataset, of 20,000  $160 \times 160$ patches sampled from a model of a subsurface binary domain. These models are assumed to have the same properties as a texture, mainly the property of global ergodicity of the data.
        \end{description}

    To avoid learning explicit pairing of real images seen by the discrimination function with constraint maps provided to the generative network, we split each dataset into training, validation and test sets, to which we add a set composed of constraint maps that should remain unrelated to the three others.
    In order to do so, a fifth of each set is used to generate the constrained pixel map $y$ by randomly selecting $0.5\%$ of the pixels from a uniform distribution, composing a set of constraints for each of the train, test and validation sets. The images from which these maps are sampled are then removed from the training, testing and validation sets. For each carried experiment the best model is selected based on some performance measures (see Section \ref{subs:eval}) computed on the validation set, as in the standard of machine learning methodology \cite{oneto2019}. Finally, reported results are computed on the test set.

        %To avoid learning explicit correlations between real examples presented to the discriminator and constraint maps given to the generator, we create a splitting consisting in the classical train, validation and test databases, to which we add a constraints database that should remain unrelated to the three others. A fifth of each set is used to generated the matrix of constraints $C$ by randomly selecting $0.5\%$ of the pixels, uniformly. These images are then removed from the training, testing and validation sets.

    \subsection{Network architectures}
	\label{subs:architectures}
	
	    We use a variety of GAN architectures in order to adapt to the different scales and image sizes of our datasets. The detailed configuration of these architectures are exposed in Appendix \ref{app:det_archis}.
	
	    For the experiments on the FashionMNIST \cite{Xiao2017}, we use a lightweight network for both the discriminator and the generator similarly to DCGAN  \cite{Radford2015} due to the small resolution of FashionMnist images.
	    %This is motivated by the large number of experiments and the small dimension of the images.
	    
	    To experiment on the Texture dataset, we consider a set of fully-convolutional generator architectures based on either dilated convolutions \cite{yu2015multi}, which behave well on texture datasets \cite{ruffino2018dilated}, or encoder-decoder architectures that are commonly used in domain-transfer applications such as CycleGAN \cite{Zhu2017unpaired}. We selected these architectures because they have very large receptive fields without using pooling, which allow the generator to use a large context for each pixel.
	    
	    We keep the same discriminator across all the experiments with these architectures, the PatchGAN discriminator \cite{Isola2017}, which is a five-layer fully-convolutional network with a sigmoid activation.

	    The Up-Dil architecture consists in a set of transposed convolutions (the upscaling part), and a set of dilated convolutional layers \cite{yu2015multi}, while the Up-EncDec has an upscaling part followed by an encoder-decoder section with skip-connections, where the constraints are downscaled, concatenated to the noise, and re-upscaled to the output size.
	    
	    The UNet \cite{ronneberger2015u} architecture is an encoder-decoder where skip-connections are added between the encoder and the decoder.
	    The Res architecture is an encoder-decoder where residual blocks \cite{he2016deep} are added after the noise is concatenated to the features. The UNet-Res combines the UNet and the Res architectures by including both residual blocks and skip-connections.
	    
	    Finally, we will evaluate our approach on the Subsurface dataset using the architecture that yields to the best performances on the Texture dataset.
	\subsection{Evaluation}
        \label{subs:eval}
        We evaluate our approach based on both the satisfaction of the pixel constraints and the visual quality of sampled images. From the assumption of Gaussian measurement noise (as discussed in Section \ref{sec:our-approach}), we assess the constraint fulfillment using the following mean square error (MSE) 
        \begin{equation}
        MSE = \frac{1}{L} \sum_{i=1}^L \left\|y_i - M(y_i) \odot G(y_i, z_i)\right\|_F^2
        \end{equation}
        This metric should be understood as the mean squared error of reconstructing the constrained pixel values. 
        
        %On one hand, we simply use the mean squared error between the provided constrained values and the constrained pixels in the generated image to evaluate the respect of the constraints.
        %On the other hand, 
        Visual quality evaluation of an image is not a trivial task \cite{Theis2015}. However, Fréchet Inception Distance (FID) \cite{Heusel2017} and Inception Score \cite{Salimans2016}, have been used to evaluate the performance of generative models. We employ FID since the Inception Score has been shown to be less reliable \cite{Barratt}. The FID consists in computing a distance between the distributions of relevant features extracted from generated and real samples. To extract these features, a pre-trained Inception v3 \cite{Szegedy2016} classifier is used to compute the embeddings of the images  at a chosen layer. Assuming these embeddings shall follow a normal distribution, the quality of the generated images is assessed in term of a Wasserstein-2 distance between the distribution of real samples and generated ones. Hence the FID writes
        %
        %It then assumes that these features are normal, and compare the (normal) distributions of the features from the real data and the fake ones using a Fréchet (or Wasserstein-2) distance, 
    
        \begin{equation}
            FID = ||\mu_r - \mu_g||^2+Tr(\Sigma_r+\Sigma_g - 2(\Sigma_r\Sigma_g)^{1/2}),
            \label{eq:fid}
        \end{equation}

        \noindent where $Tr$ is the trace operator, ($\mu_r$, $\Sigma_r$) and ($\mu_g$, $\Sigma_g$) are the pairs of mean vector and covariance matrice of embeddings obtained on respectively the real and the generated data. Being a distance between distributions,  a small FID corresponds to a good matching of the distributions.
        %corresponds to a , better the distr
        
        Since the FID requires a pre-trained classifier adapted to the dataset in study, we trained simple convolutional neural networks as classifiers for the FashionMNIST and the CIFAR-10 datasets. For the Texture dataset, since the dataset is not labeled, we resort to a CNN classifier trained on the Describable Textures Dataset (DTD) \cite{cimpoi14describing}, which is a related application domain.
        
        However, since we do not have labels for the Subsurface dataset, we could not train a classifier for this dataset, thus we cannot compute the FID. To evaluate the quality of the generated samples, we use metrics based on a distance between feature descriptors extracted from real samples and generated ones. Similarly to \cite{ruffino2018dilated}, we rely on a $\chi^2$ distance between the Histograms of Oriented Gradients (HOG) or Local Binary Patterns (LBP) features computed on generated and real images. 
        
        Histograms of Oriented Gradients (HOG) \cite{Dalal} and Local Binary Patterns (LBP) \cite{Pietikainen2011} are computed by splitting an image into cells of a given radius and computing on each cell the histograms of the oriented gradients for HOGs and of the light level differences for each pixel to the center of the cell for LBPs.  Additionally, we consider the domain-specific metric, the connectivity function \cite{lemmens2017} which is presented in Appendix \ref{app:geostatistics}.
        
        Finally, we check by visual inspection if the trained model $G$ is able to generate diverse samples, meaning that for a given $y$ and for a set of latent codes $(z_1, ..., z_n) \sim p_Z$, the generated samples $G(y,z_1), \ldots, G(y, z_n)$ are visually different. 
        
        %Since we empirically observed that our models were either producing very different samples or samples that only differ by a small noise factor, we do not propose a specific evaluation metric and instead check manually if a loss of diversity occurs.

\section{Experimental results}\label{sec:results}

    \subsection{Quality-fidelity trade-off}
    
        % \begin{figure}[!]
        %     \centering
        %     \includegraphics[trim=0 0 0 40, clip,scale=0.35]{MSE_mnist}\includegraphics[trim=0 0 0 40, clip,scale=0.35]{MSE_fashion}
        %     \includegraphics[trim=0 0 0 40, clip,scale=0.35]{FID_mnist}\includegraphics[trim=0 0 0 40, clip,scale=0.35]{FID_fashion}
            
        %     \centering
        %     \caption{MSE (top) and  FID (bottom) w.r.t. the regularization parameter $\lambda$;
        %     Dataset MNIST (left), Fashion MNIST (right).
        %     %The different orders of magnitude for the Y-axis of the FID is due to the different classifiers used to compute this distances.
        %     }
        %     \label{fig:fids}
        %     \label{fig:mses}
        % \end{figure}
        
        \begin{figure}[t]
            \centering
            \includegraphics[trim=0 0 0 40, clip,scale=0.4]{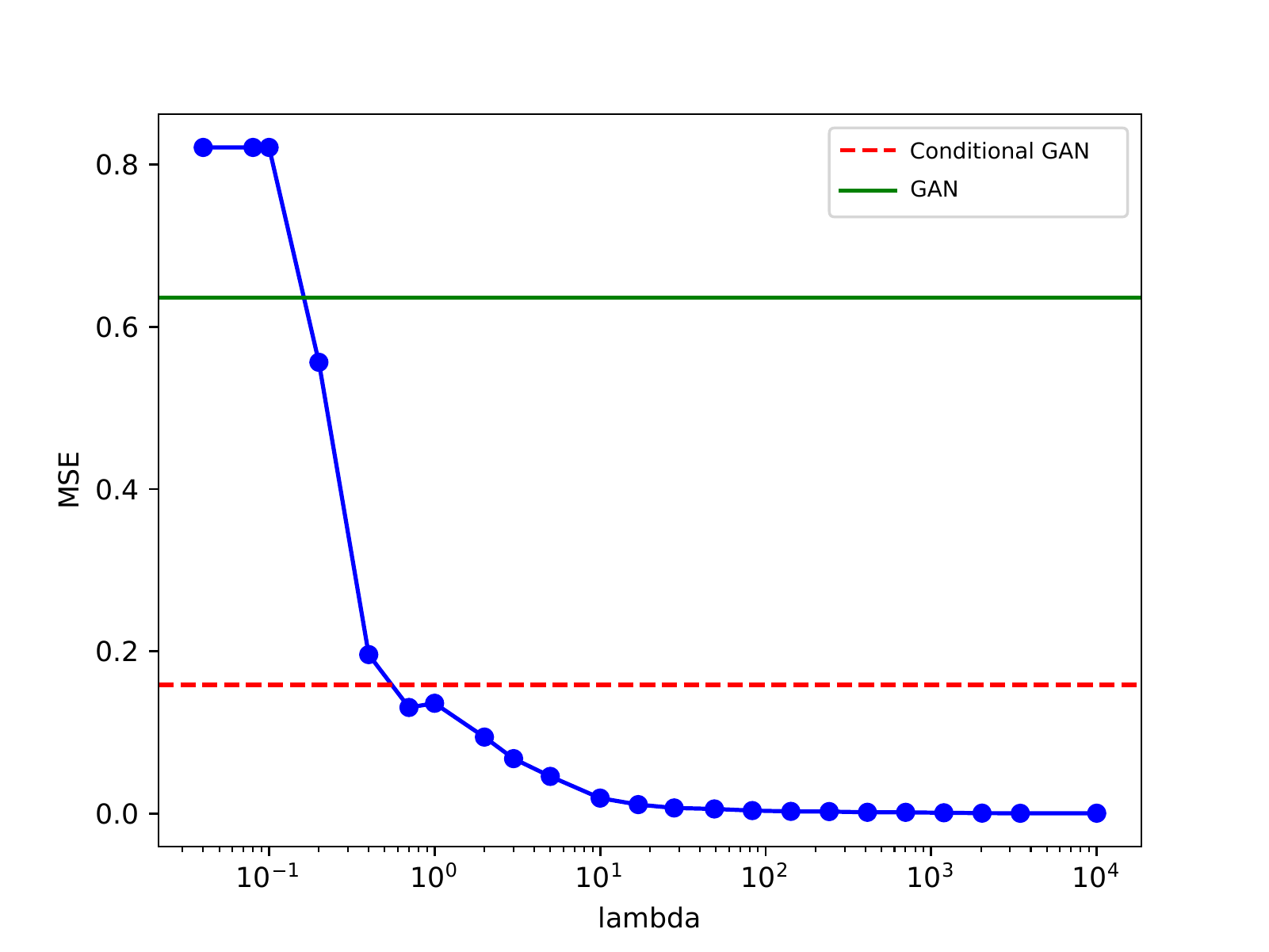}\includegraphics[trim=0 0 0 40, clip,scale=0.4]{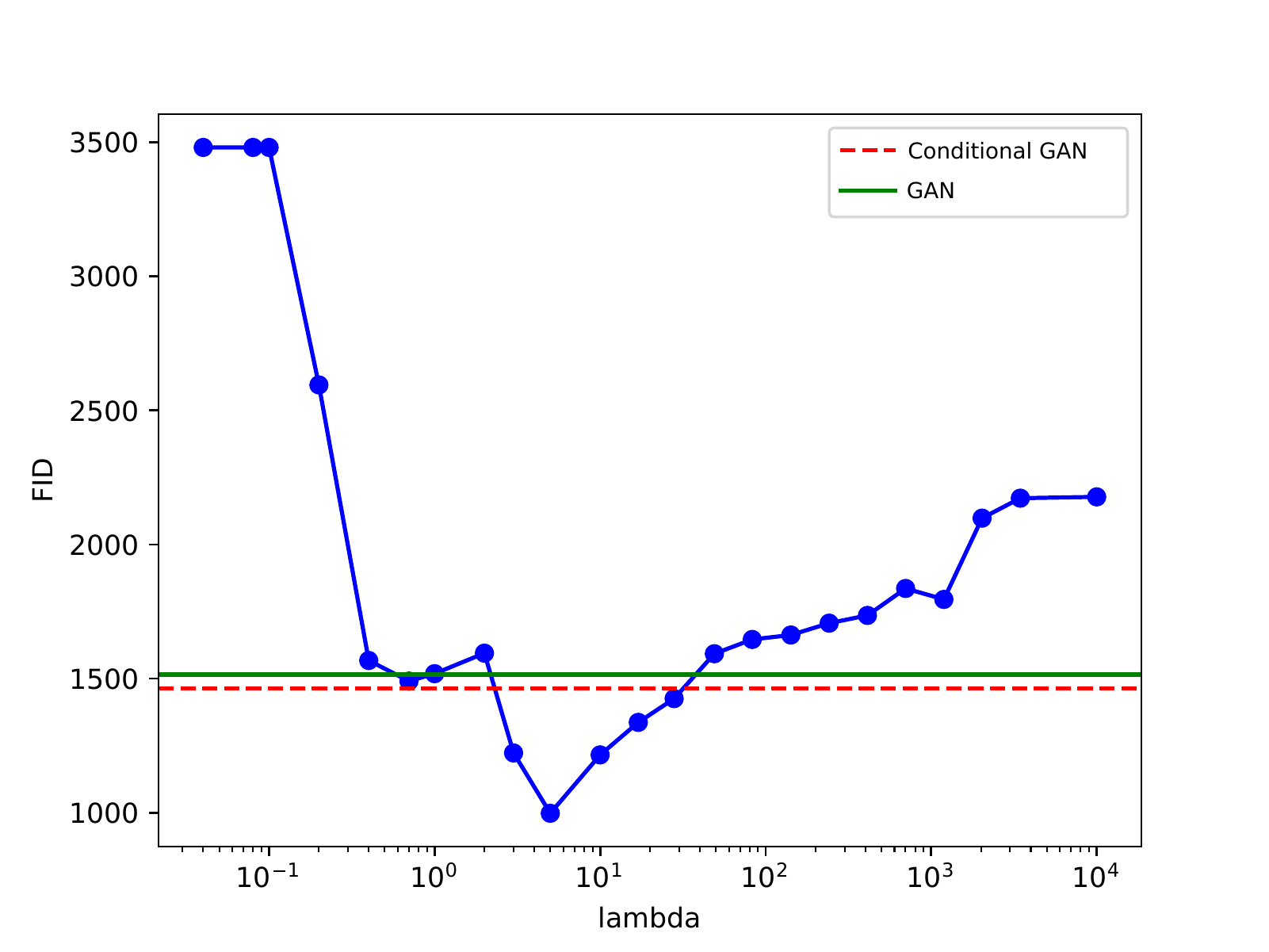}
            
            \includegraphics[scale=0.5]{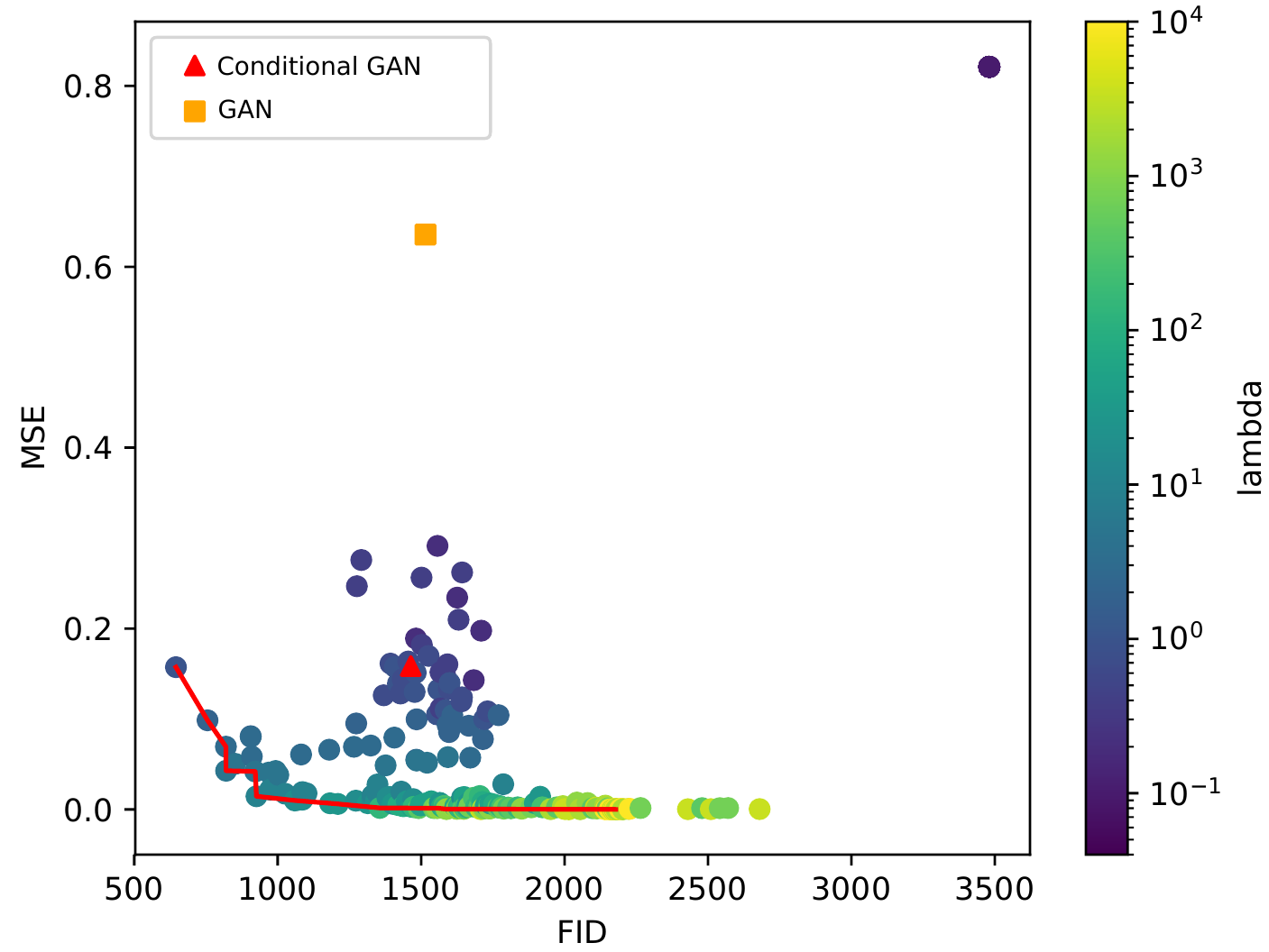}
            
            \centering
            \caption{Our approach compared to the GAN and CGAN baselines. MSE (left) and  FID (right) w.r.t. the regularization parameter $\lambda$, MSE w.r.t the FID (bottom).
            %The different orders of magnitude for the Y-axis of the FID is due to the different classifiers used to compute this distances.
            }
            \label{fig:fids}
            \label{fig:mses}
            \label{fig:paretos}
        \end{figure}
         
        %In this set of experiments, 
        We first study the influence of the $\lambda$ regularization hyper-parameter on both the quality of the generated samples and the respect of the constraints. We experiment on the %MNIST \cite{Lecun1998} and
        FashionMNIST \cite{Xiao2017} dataset, since such a study requires intensive simulations permitted by the low resolution of FashionMnist images and the used architectures (see Section \ref{subs:architectures}). 
        %a lot of re-training and the small size of the images allowed us to run several hundreds of experiments.
       
        To overcome classical GANs instability, the networks are trained 10 times and the median values of the best scores on the test set at the best epoch 
        are recorded. The epoch that minimizes:
        \begin{equation*}
            \sqrt{\left(\frac{FID - FID_{min}}{FID_{max}- FID_{min}}\right)^2 + \left(\frac{MSE - MSE_{min}}{MSE_{max}- MSE_{min}}\right)^2}
        \end{equation*}  on the validation set is considered as the best epoch, where $FID_{min}$, $MSE_{min}$, $FID_{max}$ and $MSE_{max}$ are respectively the lowest and highest FIDs and MSEs obtained on the validation set.

        Empirical evidences (highlighted in Figure \ref{fig:mses}) show that with a good choice of $\lambda$, the regularization term helps the generator to enforce the constraints, leading to smaller MSEs than when using the CGAN ($\lambda=0$) without compromising on the quality of generated images. Also, we can note that using the regularization term even leads to a better image quality compared to GAN and CGAN.
        The bottom panel in Figure \ref{fig:paretos} illustrates that the trade-off between image quality and the satisfaction of the constraints can be controlled by appropriately setting the value of $\lambda$. Nevertheless, for small values of $\lambda$ (less or equal to $10^{-1}$), our GAN model fails to learn meaningful distribution of the training images and only generates uniformly black images. This leads to the plateaus on the MSE and FID plots (top panels in Figure \ref{fig:mses}).

        % \begin{figure}
        %     \centering
        %     \includegraphics[trim=0 0 0 40, clip,scale=0.4]{pareto_mnist}\includegraphics[trim=0 0 0 40, clip,scale=0.4]{pareto_fashion}
        %     \vspace*{-3mm}
        %     \caption{MSE w.r.t the FID. Left: MNIST; Right: Fashion MNIST. Note that due to the failure mode previously mentioned, a large part of the values are stacked in the top right corner of these figures.
        %     }
        %     \label{fig:paretos}
        % \end{figure}  
      
    \subsection{Texture generation with fully-convolutional architectures}
        \label{sub:fcnn}
        Fully-convolutional architectures for GANs are widely used, either for domain-transfer applications \cite{Zhu2017unpaired}\cite{Isola2017} or for texture generation \cite{jetchev2016texture}. In order to evaluate the efficiency of our method on relatively high resolution images, we experiment the fully-convolutional networks described in Section \ref{subs:architectures} on a texture generation task using Texture dataset. We investigate the upscaling-dilatation network, the encoder-decoder one and the resnet-like architectures.
        
        Our training algorithm was run for 40 epochs on all reported results. We provide a comparison to CGAN\cite{mirza2014} approach by using the selected best architectures.
        The models are evaluated in terms of best FID (visual quality of sampled images) at each epoch and MSE (conditioning on fixed pixel values).  We also compute the FID score of the models at the epochs where the MSE is the lowest. In the other way around, the MSE is reported at epoch when the FID is the lowest. The obtained quantitative results are detailed in Table \ref{tab:ablation}.

        For the encoder-decoder models, we can notice that the models using ResNet blocks perform better than just using a UNet generator. A trade-off can also be seen between the FID and MSE for the ResNet models and the UNet-ResNet, which could mean that skip-connections help the generator to fulfill the constraints but at the price of lowered visual quality.
        
\begin{figure}
            \centering
            \includegraphics[width=2cm]{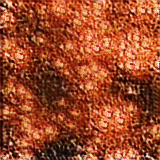}\hspace{0.5cm}\includegraphics[width=2cm]{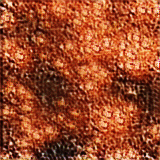}\hspace{0.5cm}\includegraphics[width=2cm]{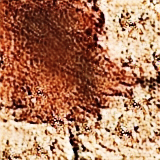}\hspace{0.5cm}\includegraphics[width=2cm]{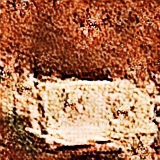}
            
            \vspace{0.3cm}
            \includegraphics[height=4cm]{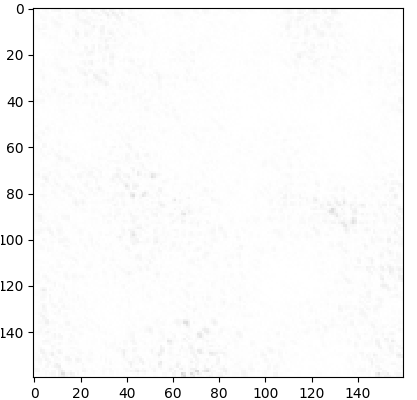}\hspace{0.5cm}\includegraphics[height=4cm]{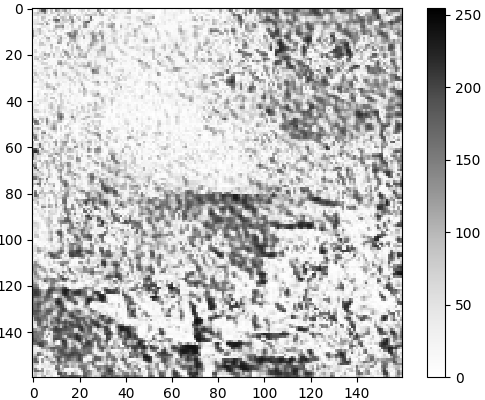}
            \caption{An example of a loss of diversity when generating Texture samples with a trained UNetRes network using two different random noises $z$ and a single constraint map $y$. The two samples on the top left are generated using the classical GAN discriminator whereas the samples on the top right are generated using the PacGAN approach. The loss of diversity is clearly visible on the absolute differences between the greyscaled images (bottom).}
            \label{fig:diversity_loss}
        \end{figure}

      Although the encoder-decoder models perform the best, they tend to lose diversity in the generated samples (see Figure \ref{fig:diversity_loss}), whereas the upscaling-based models have high FID and MSE but naturally preserve diversity in the generated samples.
        
        Changing the discriminator for a PacGAN discriminator with 2 samples in the encoder-decoder based architectures allows to restore diversity, while keeping the same performances as previously or even increasing the performances for the UNetRes (see Table \ref{tab:ablation}).
        
        Table \ref{tab:ablation-cgan} compares our proposed approach to CGAN using fully convolutional networks. It shows that our approach is more able to comply with the pixel constraints while producing realistic images. Indeed, our approach outperforms CGAN (see Table \ref{tab:ablation-cgan}) by a large margin on the respect of conditioning pixels (see the achieved MSE metrics by  our UNetPAC or UNetResPAC)  and gets  close FID performance on the generated samples. This finding is in accordance of the obtained results on FashionMnist experiments.
        %show that the comparison with the CGAN approach still holds well in a fully-convolutional setting since our approach outperforms CGAN by a large margin on the respect of the constraints and come close to it on the visual quality of the generated samples. This conforms the results obtained on the previous experiments on the FashionMNIST dataset.

    \begin{table}
        \centering
        \begin{tabular}{|l|c|c|c|c|c|}
            \hline
            Model           & Best FID & Best MSE & FID at & MSE at & Diversity\\
            &&&best MSE & best FID & \\
            \hline
            Up-Dil      & 0.0949 & 0.4137 & 1.0360 & 0.7057 & {\color{green}\cmark } \\
            Up-EncDec  & 0.1509 & 0.7570 & 0.2498 & 0.9809 & {\color{green}\cmark } \\
            UNet        & 0.0442 & 0.1789 & 0.0964 & 0.4559 & {\color{red}\xmark } \\
            Res      & 0.0458 & 0.0474 & 0.0590 & 0.0476 & {\color{red}\xmark } \\
            UNetRes & 0.0382 & 0.0307 & 0.0499 & 0.0338 & {\color{red}\xmark } \\
            \hline
            ResPAC &  \textbf{0.0350} & 0.0698 & 0.0466 & 0.4896 & {\color{green}\cmark } \\
            UNetPAC &  0.0672 & \textbf{$\leq$ 0.0001} & 0.3120 & 0.2171&  {\color{green}\cmark } \\
            UNetResPAC & 0.0431 & 0.0277 & \textbf{0.0447} & \textbf{0.0302} &  {\color{green}\cmark }\\
            \hline
        \end{tabular}
        
        \caption{Results obtained by the different fully-convolutional architectures on the Texture dataset. We can remark that the encoder-decoder greatly outperforms the upscaling ones and that using the PacGAN technique helps keeping the performance of these models while restoring the diversity in the samples. The bottom part of the table refers to PacGan architectures.}
        \label{tab:ablation}
    \end{table}
    
       \begin{table}[t]
        \centering
        \begin{tabular}{|l|c|c|c|c|c|}
                 \hline
            Model           & Best FID & Best MSE & FID at & MSE at \\
            &&&best MSE & best FID  \\
            \hline
            CGAN-ResPAC &   \textbf{0.0234} & 0.1337 &  \textbf{0.0340} & 0.2951 \\
            CGAN-UNetPAC &  0.0518 & 0.2010 & 0.0705 & 0.4828\\
            CGAN-UNetResPAC & 0.0428 & 0.1060 & 0.0586 & 0.2250\\
            \hline
            Ours-ResPAC &  0.0350 & 0.0698 & 0.0466 & 0.4896\\
            Ours-UNetPAC &  0.0672 & \textbf{$\leq$ 0.0001}  & 0.3120 & 0.2171 \\
            Ours-UNetResPAC & 0.0431 & 0.0277 &0.0447 & \textbf{0.0302}\\
            \hline
        \end{tabular}
        
        \caption{Results obtained by the selected best fully-convolutional architectures on the Texture dataset for both the CGAN approach and our approach.}
        \label{tab:ablation-cgan}
    \end{table}
    
    \subsection{Extended architectures}
        We extend the comparison of our approach to CGAN on the CIFAR10 and CelebA  datasets (Table \ref{tab:cifar10}). We investigated the architectures described in Section \ref{subs:architectures}. All reported results are obtained with the regularization parameter fixed to $\lambda=1$.
        We train the networks for 150 epochs using the same dataset split as stated previously in order to keep independence between the images constraint maps. The evaluation procedure remains also unchanged. We use the PacGAN approach to avoid the loss of diversity issues. The experiments on both datasets show that though CGAN  provides better results in terms of visual quality, our approach outperforms it according to the respect of the pixel constraints.

    \begin{table}
        \centering
        \begin{tabular}{|l|c|c|c|c|c|}
            \hline
            &Model           & Best FID & Best MSE & FID at & MSE at \\
            &&&&best MSE & best FID \\
            \hline
            CIFAR-10 &CGAN   & \textbf{2,68}  & 0.081  & \textbf{2.68}  & 0.081\\
            &Ours            & 3.120 & \textbf{0.010} & 3.530 & \textbf{0.011} \\    
            \hline
            CelebA &CGAN      & \textbf{1.34e-4} & 0.0209 &  \textbf{1.81e-4} & 0.0450\\
            &Ours            & 2.09e-4& \textbf{0.0053} & 5.392e-4 & \textbf{0.0249} \\
            \hline
        \end{tabular}
        
        \caption{Results on the CIFAR10 and CelebA datasets. The reported performances compare CGAN to our proposed GAN conditioned on scarce constraint map.}
        \label{tab:cifar10}
    \end{table}
    
    \vspace{0.4cm}
    
    \subsection{Application to hydro-geology}
    
        Finally, we evaluate our approach on the Subsurface dataset. We use the UNetResPAC  architecture, since it performed the best on Texture data as exposed in Section \ref{sub:fcnn}. As previously, we simply set the regularization parameter at $\lambda=1$ and, the network is trained for 40 epochs using the same experimental protocol. To evaluate the trade-off between the visual quality and the respect of the constraints, instead of FID we rather compute distances between visual Histograms of Oriented Gradients (see Section \ref{sec:experiments_protocol}), extracted from real and generated samples. We also evaluate the visual quality of our approach with a distance between Local Binary Patterns. Indeed, Subsurface application lacks labelled data in order to learn a deep network classifier from which the FID score can be computed. 
        
        %As stated before in Section \ref{subs:eval}, we cannot use the FID to evaluate the visual quality of the generated images since we don't have a supervised task linked to the data.
        %Therefore we use distances between visual features, namely Histograms of Oriented Gradients and Local Binary Patterns (see Section \ref{sec:experiments_protocol}), extracted from real and generated samples.
        
        The obtained results are summarized in Tables \ref{tab:subsurface} and \ref{tab:subsurface_visual}. They are coherent with the previous experiments since the generated samples are diverse and have a low error regarding the constrained pixels. The conditioning have a limited impact on the visual quality of the generated samples and compares well to unconditional approaches \cite{ruffino2018dilated}. Evaluation of the generated images using the domain-connectivity function highlights this fact on Figures \ref{fig:ours_connectivity} and \ref{fig:ours_connectivity} in the supplementary materials. Also examples of generated images by our approach  pictured in Figure \ref{fig:samples_subsurface} (see appendix \ref{app:generated_images}) show that we preserve the visual quality and honor the constraints.
        
    \begin{table}
        \centering
        \begin{tabular}{|l|c|c|c|c|c|}
            \hline
        &Model           & Best HOG & Best MSE& HOG at & MSE at \\
            &&& &  best MSE & best HOG \\
            \hline
            Subsurface &CGAN   & \textbf{2.92e-4} & 0.2505 & \textbf{3.06e-4}  & 1.1550 \\
            &Ours            & 4.31e-4 & \textbf{0.0325}& 5.69e-4 & \textbf{0.2853} \\
            \hline
          \end{tabular}
        \caption{Evaluation of the trade-off between the visual quality of the generated samples and the respect of the constraints for the CGAN approach and ours on the Subsurface dataset.}
        \label{tab:subsurface}
    \end{table}

    \begin{table}[h]
        \centering
        \begin{tabular}{|l|c|c|c|c|c|}
            \hline
            &Model           & Best HOG & Best MSE& Best LBP & Best LBP \\
            &&& & (radius=1) & (radius=2) \\
            \hline
            Subsurface &CGAN   & \textbf{2.92e-4} & 0.2505 & \textbf{2.157} & \textbf{3.494}\\
            &Ours            &  4.31e-4 &\textbf{0.0325} & 10.142 & 16.754 \\
            \hline
        \end{tabular}
        \caption{Evaluation of the visual quality between the CGAN approach and ours on the Subsurface dataset using several metrics.}
        \label{tab:subsurface_visual}
    \end{table}

\section*{Conclusion}
In this paper, we address the task of learning effective generative adversarial networks when only very few pixel values are known beforehand. To solve this pixel-wise conditioned GAN, we model the conditioning information under a probabilistic framework. This leads to the maximization of the likelihood of the constraints given a
generated image. Under the assumption of a Gaussian distribution over the given pixels, we formulate an objective function composed of the conditional GAN loss function regularized by a $\ell_2$-norm on pixel reconstruction errors. We describe the related optimization algorithm.

Empirical evidences illustrate that the proposed framework helps obtaining good image quality while best fulfilling the constraints compared to classical GAN approaches. We show that, if we include the PacGAN technique,  this  approach  is  compatible  with  fully-convolutional  architectures  and scales well to large images. We apply this approach to a common geological simulation task and show that it allows the generation of realistic samples which fulfill the prescribed constraints.

In future work, we plan to investigate other prior distributions for the given pixels as the Laplacian or $\beta$-distribtutions. We are also interested in applying the developed approach to other applications or signals such as audio inpainting \cite{marafioti2018context}.

\section*{Acknowledgements}
This research was supported by the CNRS PEPS I3A REGGAN project and the ANR-16-CE23-0006 grant \emph{Deep in France}. We kindly thank the CRIANN for the provided high-computation facilities. 
%who put the computational resources used for this paper to our disposition.
	
% ****************************************************************************
% BIBLIOGRAPHY AREA
% ****************************************************************************

\begin{footnotesize}

\bibliographystyle{unsrt}
\bibliography{neucom}

\end{footnotesize}

\newpage

\begin{appendices}

\section{Details of the datasets}
\label{app:det_datasets}

  {
        \centering
        \begin{tabular}{|l|c|c|c|c|c|}
            \hline
            Dataset           & Size (in pixels)& Training set & Validation set & Test set\\
            \hline
            FashionMNIST &28x28& 55,000 & 5,000 & 10,000\\
            Cifar-10 & 32x32& 55,000 & 5,000 & 10,000 \\
            CelebA & 128x128 & 80,000 & 5,000 & 15,000\\
            \hline
            Texture & 160x160 & 20,000 & 2,000 & 4,000\\
            Subsurface & 160x160 & 20,000 & 2,000 & 4,000\\
            \hline
        \end{tabular}
}\\

Additional information: \begin{itemize}
    \item For FashionMNIST and Cifar-10, we keep the original train/test split and then sample 5000 images from the training set that act as validation samples.
    \item For the Texture dataset, we sample patches randomly from a 3840x2400 image of a brick wall.

\end{itemize}

\section{Detailed deep architectures}
\label{app:det_archis}
 
 \subsection{DCGAN for FashionMNIST}
 {
        \centering
        \begin{tabular}{|l|c|c|c|c|c|}
            \hline
            Layer type & Units & Scaling & Activation & Output shape\\
            \hline
            Input z & - & - & - & 7x7\\
            Input y & - & - & - & 28x28\\
            Dense & 343 & - & ReLU & 7x7\\
            Conv2DTranspose & 128 3x3 & x2 & ReLU & 14x14 \\
            Conv2DTranspose & 64 3x3 & x2 & ReLU & 28x28 \\
            Conv2DTranspose & 1 3x3 & x1 & tanh & 28x28 \\
            \hline
            Input x & - & - & - & 28x28\\
            Input y & - & - & - & 28x28\\
            Conv2D & 64 3x3 & x1/2 & LeakyReLU & 14x14 \\
            Conv2D & 128 3x3 & x1/2 & LeakyReLU & 7x7 \\
            Conv2D & 1 3x3 & x1 & tanh & 28x28 \\
            Dense & 1 & - & Sigmoid & 1\\
            \hline
        \end{tabular}
}\\ 

Additional information: \begin{itemize}
    \item Batch normalization\cite{ioffe2015batch} is applied across all the layers
    \item A Gaussian noise is applied to the input of the discriminator
\end{itemize}

\subsection{UNet-Res for CIFAR10} \label{subsec:Unet-Cifar}
 {
        \centering
        \begin{tabular}{|l|c|c|c|c|c|}
            \hline
            Layer type & Units & Scaling & Activation & Output shape\\
            \hline
            Input y & - & - & - & 32x32\\
            Conv2D* & 64 5x5 & x1 & ReLU & 32x32 \\
            Conv2D* & 128 3x3 & x1/2 & ReLU & 16x16 \\
            Conv2D* & 256 3x3 & x1/2 & ReLU & 8x8 \\
            Input z & - & - & - & 8x8\\
            Dense & 256 & - & ReLU & 8x8\\
            Residual block & 3x256 3x3 & x1 & ReLU & 8x8 \\
            Residual block & 3x256 3x3 & x1 & ReLU & 8x8 \\
            Residual block & 3x256 3x3 & x1 & ReLU & 8x8 \\
            Residual block & 3x256 3x3 & x1 & ReLU & 8x8 \\
            Conv2DTranspose* & 256 3x3 & x2 & ReLU & 16x16 \\
            Conv2DTranspose* & 128 3x3 & x2 & ReLU & 32x32 \\
            Conv2DTranspose* & 64 3x3 & x1 & ReLU & 32x32 \\
            Conv2D & 3 3x3 & x1 & tanh & 32x32 \\
            \hline
            Input x & - & - & - & 32x32\\
            Input y & - & - & - & 32x32\\
            Conv2D & 64 3x3 & x1/2 & LeakyReLU & 16x16 \\
            Conv2D & 128 3x3 & x1/2 & LeakyReLU & 8x8 \\
            Conv2D & 256 3x3 & x1/2 & LeakyReLU & 4x4 \\
            Dense & 1 & - & Sigmoid & 1\\
            \hline
        \end{tabular}
}\\ 

Additional information: \begin{itemize}
    \item Instance normalization\cite{ulyanov2016instance} is applied across all the layers instead of Batch normalization. This is involved by the use of the PacGAN technique.
    \item A Gaussian noise is applied to the input of the discriminator
    \item The layers noted with an asterisk are linked with a skip-connection
\end{itemize}

\subsection{UNet-Res for CelebA}
\label{subsec:unet_celeba}
 {
        \centering
        \begin{tabular}{|l|c|c|c|c|c|}
            \hline
            Layer type & Units & Scaling & Activation & Output shape\\
            \hline
            Input y & - & - & - & 128x128\\
            Conv2D & 64 5x5 & x1 & ReLU & 128x128 \\
            Conv2D* & 128 3x3 & x1/2 & ReLU & 64x64 \\
            Conv2D* & 256 3x3 & x1/2 & ReLU & 32x32 \\
            Conv2D* & 512 3x3 & x1/2 & ReLU & 16x16 \\
            Input z & - & - & - & 16x16\\
            Dense & 256 & - & ReLU & 16x16\\
            Residual block & 3x256 3x3 & x1 & ReLU & 16x16 \\
            Residual block & 3x256 3x3 & x1 & ReLU & 16x16 \\
            Residual block & 3x256 3x3 & x1 & ReLU & 16x16 \\
            Residual block & 3x256 3x3 & x1 & ReLU & 16x16 \\
            Residual block & 3x256 3x3 & x1 & ReLU & 16x16 \\
            Residual block & 3x256 3x3 & x1 & ReLU & 16x16 \\
            Conv2DTranspose* & 256 3x3 & x2 & ReLU & 32x32 \\
            Conv2DTranspose* & 128 3x3 & x2 & ReLU & 64x64 \\
            Conv2DTranspose* & 64 5x5 & x2 & ReLU & 128x128 \\
            Conv2D & 3 3x3 & x1 & tanh & 128x128 \\
            \hline
            Input x & - & - & - & 128x128\\
            Input y & - & - & - & 128x128\\
            Conv2D & 64 3x3 & x1/2 & LeakyReLU & 64x64 \\
            Conv2D & 128 3x3 & x1/2 & LeakyReLU & 32x32 \\
            Conv2D & 256 3x3 & x1/2 & LeakyReLU & 16x16 \\
            Conv2D & 512 3x3 & x1/2 & LeakyReLU & 32x32 \\
            Dense & 1 & - & Sigmoid & 1\\
            \hline
        \end{tabular}
}\\~\\
\noindent
This network follows the same additional setup as described in Appendix (\ref{subsec:Unet-Cifar}).
%Additional details: \begin{itemize}
 %   \item Instance normalization\cite{ulyanov2016instance} is applied across all the layers instead of Batch normalization. This is due to the use of the PacGAN technique.
  %  \item Gaussian noise is applied to the input of the discriminator
   % \item The layers noted with an asterisked are linked with a skip-connection
%\end{itemize}

\subsection{Architectures for Texture}

\subsubsection{PatchGAN discriminator}
 {
        \centering
        \begin{tabular}{|l|c|c|c|c|c|}
            \hline
            Layer type & Units & Scaling & Activation & Output shape\\
            \hline
            Input x & - & - & - & 160x160\\
            Input y & - & - & - & 160x160\\
            Conv2D & 64 3x3 & x1/2 & LeakyReLU & 80x80 \\
            Conv2D & 128 3x3 & x1/2 & LeakyReLU & 40x40 \\
            Conv2D & 256 3x3 & x1/2 & LeakyReLU & 20x20 \\
            Conv2D & 512 3x3 & x1/2 & LeakyReLU &  10x10\\
            \hline
        \end{tabular}
} 

\subsubsection{UpDil Texture}
 {
        \centering
        \begin{tabular}{|l|c|c|c|c|c|}
            \hline
            Layer type & Units & Scaling & Activation & Output shape\\
            \hline
            Input z & - & - & - & 20x20 \\
            Conv2DTranspose & 256 3x3 & x2 & ReLU & 40x40 \\
            Conv2DTranspose & 128 3x3 & x2 & ReLU & 80x80 \\
            Conv2DTranspose & 64 3x3 & x2 & ReLU & 160x160 \\
            Input y & - & - & - & 160x160\\
            Conv2D & 64 3x3 dil. 1 & x1 & ReLU & 160x160 \\
            Conv2D & 128 3x3  dil. 2 & x1 & ReLU & 160x160 \\
            Conv2D & 256 3x3 dil. 3 & x1 & ReLU & 160x160 \\
            Conv2D & 512 3x3  dil. 4& x1 & ReLU & 160x160 \\
            Conv2D & 3 3x3 & x1 & tanh & 160x160 \\
            \hline
        \end{tabular}
}

\subsubsection{UpEncDec Texture}
 {
        \centering
        \begin{tabular}{|l|c|c|c|c|c|}
            \hline
            Layer type & Units & Scaling & Activation & Output shape\\
            \hline
            Input z & - & - & - & 20x20 \\
            Conv2DTranspose & 256 3x3 & x2 & ReLU & 40x40 \\
            Conv2DTranspose & 128 3x3 & x2 & ReLU & 80x80 \\
            Conv2DTranspose & 64 5x5 & x2 & ReLU & 160x160 \\
            Input* y & - & - & - & 160x160\\
            Conv2D* & 64 3x3 & x1/2 & ReLU & 80x80 \\
            Conv2D* & 128 3x3 & x1/2 & ReLU & 40x40 \\
            Conv2D & 256 3x3 & x1/2 & ReLU & 20x20 \\
            Conv2DTranspose* & 256 3x3 & x2 & ReLU & 40x40 \\
            Conv2DTranspose* & 128 3x3 & x2 & ReLU & 80x80 \\
            Conv2DTranspose* & 64 3x3 & x2 & ReLU & 160x160 \\
            Conv2D & 3 3x3 & x1 & tanh & 160x160 \\
            \hline
        \end{tabular}
}

\subsubsection{UNet Texture}
 {
        \centering
        \begin{tabular}{|l|c|c|c|c|c|}
            \hline
            Layer type & Units & Scaling & Activation & Output shape\\
            \hline
            Input y & - & - & - & 160x160\\
            Conv2D & 64 5x5 & x1 & ReLU & 160x160 \\
            Conv2D* & 128 3x3 & x1/2 & ReLU & 80x80 \\
            Conv2D* & 256 3x3 & x1/2 & ReLU & 40x40 \\
            Conv2D* & 512 3x3 & x1/2 & ReLU & 20x20 \\
            Input z & - & - & - & 20x20 \\
            Conv2DTranspose* & 256 3x3 & x2 & ReLU & 40x40 \\
            Conv2DTranspose* & 128 3x3 & x2 & ReLU & 80x80 \\
            Conv2DTranspose* & 64 5x5 & x2 & ReLU & 160x160 \\
            Conv2D & 3 3x3 & x1 & tanh & 160x160 \\
            \hline
        \end{tabular}
}

\subsubsection{Res Texture}
 {
        \centering
        \begin{tabular}{|l|c|c|c|c|c|}
            \hline
            Layer type & Units & Scaling & Activation & Output shape\\
            \hline
            Input y & - & - & - & 160x160\\
            Conv2D & 64 5x5 & x1 & ReLU & 160x160 \\
            Conv2D & 128 3x3 & x1/2 & ReLU & 80x80 \\
            Conv2D & 256 3x3 & x1/2 & ReLU & 40x40 \\
            Conv2D & 512 3x3 & x1/2 & ReLU & 20x20 \\
            Input z & - & - & - & 20x20 \\
            Residual block & 3x256 3x3 & x1 & ReLU & 20x20 \\
            Residual block & 3x256 3x3 & x1 & ReLU & 20x20 \\
            Residual block & 3x256 3x3 & x1 & ReLU & 20x20 \\
            Residual block & 3x256 3x3 & x1 & ReLU & 20x20 \\
            Residual block & 3x256 3x3 & x1 & ReLU & 20x20 \\
            Residual block & 3x256 3x3 & x1 & ReLU & 20x20 \\
            Conv2DTranspose & 256 3x3 & x2 & ReLU & 40x40 \\
            Conv2DTranspose & 128 3x3 & x2 & ReLU & 80x80 \\
            Conv2DTranspose & 64 5x5 & x2 & ReLU & 160x160 \\
            Conv2D & 3 3x3 & x1 & tanh & 160x160 \\
            \hline
        \end{tabular}
} 

\subsubsection{UNet-Res Texture} 
 {
        \centering
        \begin{tabular}{|l|c|c|c|c|c|}
            \hline
            Layer type & Units & Scaling & Activation & Output shape\\
            \hline
            Input y & - & - & - & 160x160\\
            Conv2D & 64 5x5 & x1 & ReLU & 160x160 \\
            Conv2D* & 128 3x3 & x1/2 & ReLU & 80x80 \\
            Conv2D* & 256 3x3 & x1/2 & ReLU & 40x40 \\
            Conv2D* & 512 3x3 & x1/2 & ReLU & 20x20 \\
            Input z & - & - & - & 20x20 \\
            Residual block & 3x256 3x3 & x1 & ReLU & 20x20 \\
            Residual block & 3x256 3x3 & x1 & ReLU & 20x20 \\
            Residual block & 3x256 3x3 & x1 & ReLU & 20x20 \\
            Residual block & 3x256 3x3 & x1 & ReLU & 20x20 \\
            Residual block & 3x256 3x3 & x1 & ReLU & 20x20 \\
            Residual block & 3x256 3x3 & x1 & ReLU & 20x20 \\
            Conv2DTranspose* & 256 3x3 & x2 & ReLU & 40x40 \\
            Conv2DTranspose* & 128 3x3 & x2 & ReLU & 80x80 \\
            Conv2DTranspose* & 64 5x5 & x2 & ReLU & 160x160 \\
            Conv2D & 3 3x3 & x1 & tanh & 160x160 \\
            \hline
        \end{tabular}
}\\~\\
\noindent
As for Cifar10, this network follows the same additional setup described in Appendix (\ref{subsec:Unet-Cifar}).

\section{Domain-specific metrics for underground soil generation}
\label{app:geostatistics}

In this section, we compute the connectivity function \cite{lemmens2017} of generated soil image, a domain-specific metric, which is the probability that a continuous pixel path exists between two pixels of the same value (called Facies) in a given direction and a given distance (called Lag). This connectivity function should be similar to the one obtained on real-world samples. In this application, the connectivity function models the probability that two given pixels are from the same sand brick or clay matrix zone.

\begin{figure}[t]
    \centering
    \includegraphics[scale=0.45]{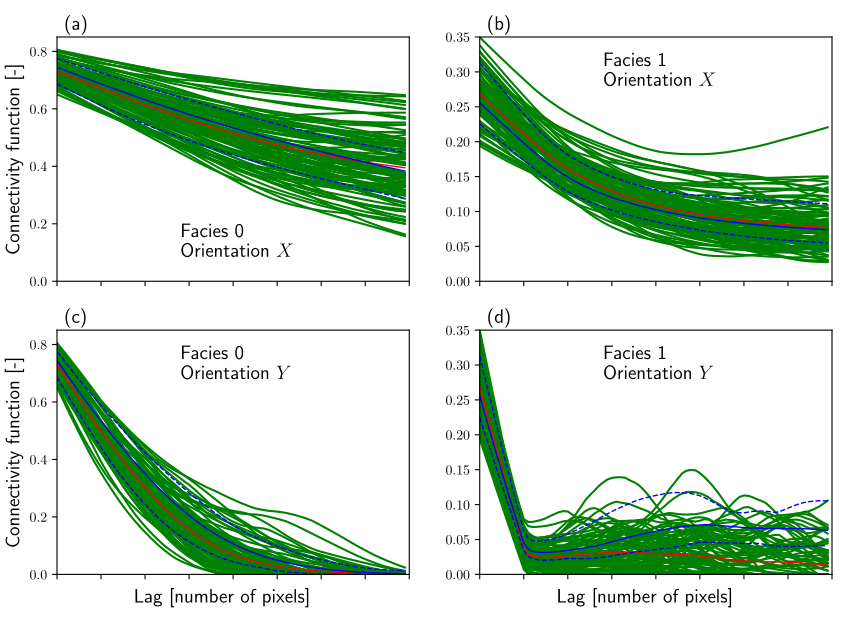}
    \caption{Connectivity curves obtained on 100 samples generated with the CGAN approach.}
    \label{fig:CGAN_connectivity}
\end{figure}

\begin{figure}[!t]
    \centering
    \includegraphics[scale=0.45]{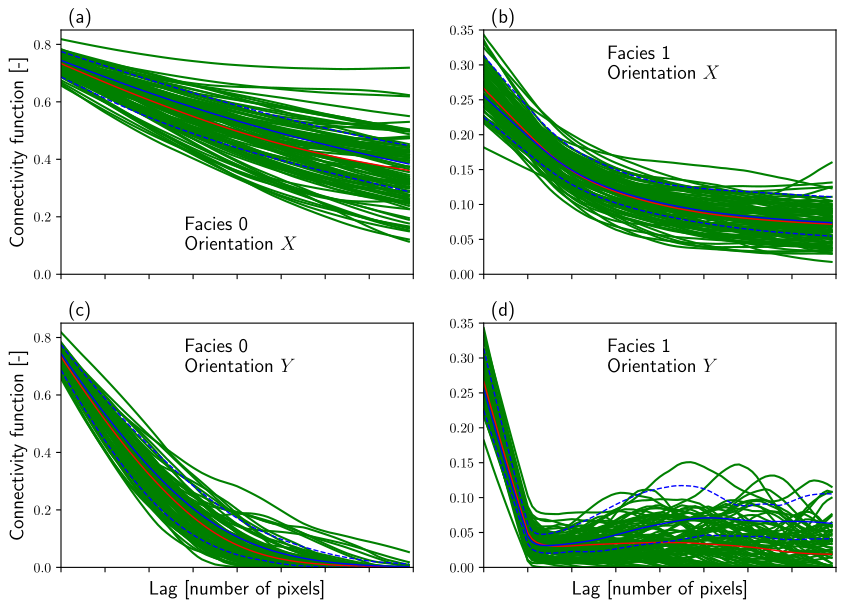}
    \caption{Connectivity curves obtained on 100 samples generated with our approach.}
    \label{fig:ours_connectivity}
\end{figure}

We sampled 100 real and 100 generated images using the UNetResPAC architecture (see Section \ref{subs:architectures}) on which the connectivity function was evaluated for both the CGAN and our approach. The obtained graphs are shown respectively in Figures \ref{fig:CGAN_connectivity} and \ref{fig:ours_connectivity}.

The blue curves are the mean value for the real samples, and the blue dashed curves are the minimum and maximum values on these samples. The green curves are the connectivity functions for each of the 100 synthetic samples and the red curves are their mean connectivity functions.
From these curves
%the Figures \ref{fig:CGAN_connectivity} and \ref{fig:ours_connectivity}, 
we observe that that our approach has similar connectivity functions as the CGAN approach while being significantly better at respecting the given constraints (see Section Table \ref{tab:subsurface}). 

\FloatBarrier

\section{Additional samples from the Texture and Subsurface datasets}
\label{app:generated_images}

\begin{figure}
    \centering
    Texture: Real samples\\
    \includegraphics[scale=0.3]{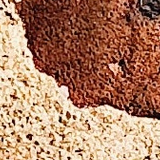}\hspace{5px}\includegraphics[scale=0.3]{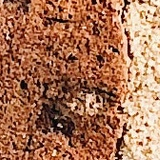}\hspace{5px}\includegraphics[scale=0.3]{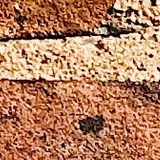}\hspace{5px}\includegraphics[scale=0.3]{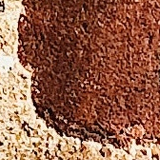}\hspace{5px}\includegraphics[scale=0.3]{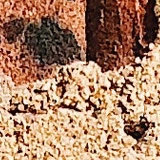}\hspace{5px}\includegraphics[scale=0.3]{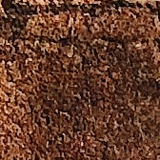}\\
    \vspace{0.1cm}
    \includegraphics[scale=0.3]{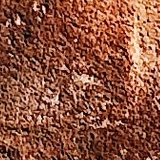}\hspace{5px}\includegraphics[scale=0.3]{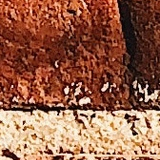}\hspace{5px}\includegraphics[scale=0.3]{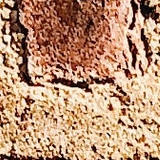}\hspace{5px}\includegraphics[scale=0.3]{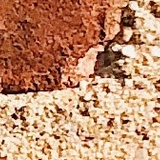}\hspace{5px}\includegraphics[scale=0.3]{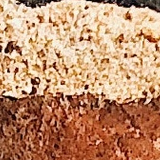}\hspace{5px}\includegraphics[scale=0.3]{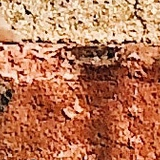}\\
    
    \dotfill
    
    Texture: Generated samples\\
    \includegraphics[scale=0.3]{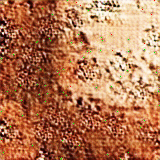}\hspace{5px}\includegraphics[scale=0.3]{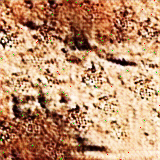}\hspace{5px}\includegraphics[scale=0.3]{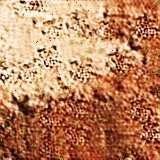}\hspace{5px}\includegraphics[scale=0.3]{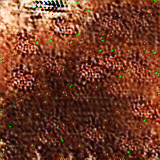}\hspace{5px}\includegraphics[scale=0.3]{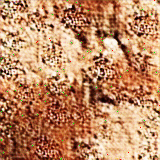}\hspace{5px}\includegraphics[scale=0.3]{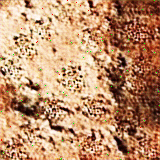}\\
    \vspace{0.1cm}
    \includegraphics[scale=0.3]{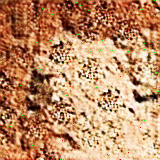}\hspace{5px}\includegraphics[scale=0.3]{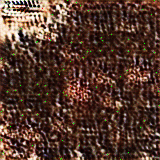}\hspace{5px}\includegraphics[scale=0.3]{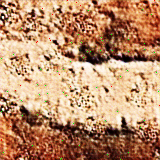}\hspace{5px}\includegraphics[scale=0.3]{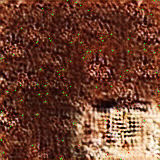}\hspace{5px}\includegraphics[scale=0.3]{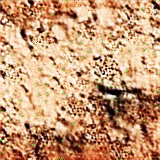}\hspace{5px}\includegraphics[scale=0.3]{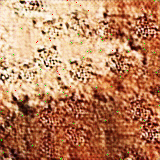}\\
    
         \caption{Real and generated samples from the Texture dataset.}
    \label{fig:samples_texture}
    \end{figure}

\begin{figure}[t]
\centering
  Subsurface: Real samples\\
    \includegraphics[scale=0.3]{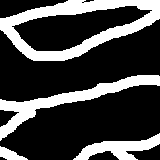}\hspace{5px}\includegraphics[scale=0.3]{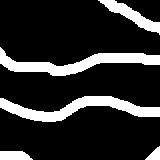}\hspace{5px}\includegraphics[scale=0.3]{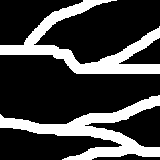}\hspace{5px}\includegraphics[scale=0.3]{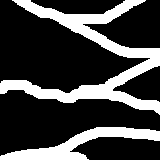}\hspace{5px}\includegraphics[scale=0.3]{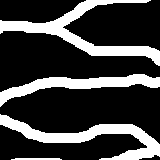}\hspace{5px}\includegraphics[scale=0.3]{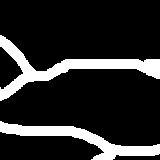}\\
    \vspace{0.1cm}
    \includegraphics[scale=0.3]{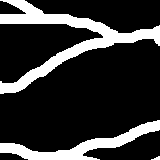}\hspace{5px}\includegraphics[scale=0.3]{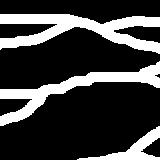}\hspace{5px}\includegraphics[scale=0.3]{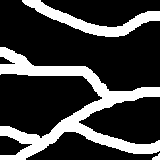}\hspace{5px}\includegraphics[scale=0.3]{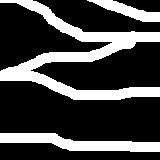}\hspace{5px}\includegraphics[scale=0.3]{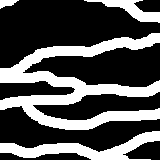}\hspace{5px}\includegraphics[scale=0.3]{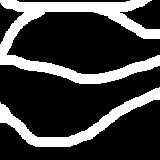}
    
    \dotfill
    
    Subsurface: Generated samples\\
    \includegraphics[scale=0.3]{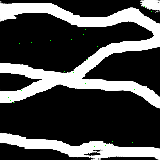}\hspace{5px}\includegraphics[scale=0.3]{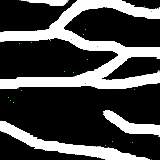}\hspace{5px}\includegraphics[scale=0.3]{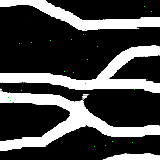}\hspace{5px}\includegraphics[scale=0.3]{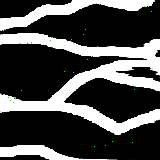}\hspace{5px}\includegraphics[scale=0.3]{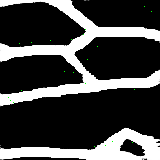}\hspace{5px}\includegraphics[scale=0.3]{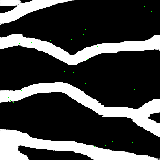}\\
    \vspace{0.1cm}
    \includegraphics[scale=0.3]{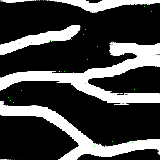}\hspace{5px}\includegraphics[scale=0.3]{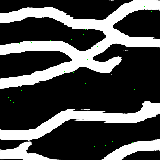}\hspace{5px}\includegraphics[scale=0.3]{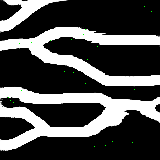}\hspace{5px}\includegraphics[scale=0.3]{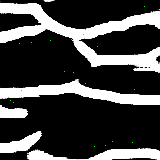}\hspace{5px}\includegraphics[scale=0.3]{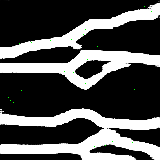}\hspace{5px}\includegraphics[scale=0.3]{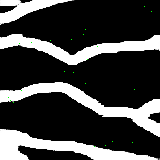}\\
    
     \caption{Real and generated samples from the Subsurface dataset.}
    \label{fig:samples_subsurface}
\end{figure}

In this section, we show some samples generated with the UNetResPAC architecture, which performs the best in our experiments (see Section \ref{sec:results}) compared to real images sampled from the Texture (Figure \ref{fig:samples_texture}) and Subsurface (Figure \ref{fig:samples_subsurface}) datasets. For the generated samples, the enforced pixel constraints are colored in the images, green corresponding to a squared error less than $0.1$ and red otherwise.


\end{appendices}

% ****************************************************************************
% END OF BIBLIOGRAPHY AREA
% ****************************************************************************

\end{document}